\documentclass[12pt]{article}
\usepackage{amsfonts}

\let\ssection=\section
\renewcommand{\section}{\setcounter{equation}{0}\ssection}

\newcommand\mathC{\mkern1mu\raise2.2pt\hbox{$\scriptscriptstyle|$}
        {\mkern-7mu\rm C}} 
\newcommand{\mathR}{{\rm I\! R}}         

\renewcommand{\o}{\ensuremath{{\omega}}}
\renewcommand{\O}{\ensuremath{{\Omega}}}

\newcommand\bi{\begin{itemize}}
\newcommand\ei{\end{itemize}}
\newcommand\be{\begin{equation}}
\newcommand\ee{\end{equation}}

\newcommand{\dd}{\ensuremath{{\delta}}}
\newcommand{\al}{\ensuremath{{\alpha}}}
\newcommand{\bb}{\ensuremath{{\beta}}}
\newcommand{\pl}{\ensuremath{{\partial}}}

\newcommand{\fl}{\ensuremath{{\flat}}}

\newcommand{\sh}{\ensuremath{{\sharp}}}

\begin{document}
\begin{titlepage}

\begin{center}

{\large\bf On Symmetry and Conserved Quantities in Classical Mechanics}
\end{center}

\vspace{0.3 truecm}

\begin{center}
        J.~Butterfield\footnote{email: jb56@cus.cam.ac.uk;
            jeremy.butterfield@all-souls.oxford.ac.uk}\\
            [10pt] All Souls College\\ Oxford OX1 4AL
\end{center}

\begin{center}
     Tuesday 12 July 2005; for a Festschrift for Jeffrey Bub, ed. W.   
                Demopoulos and I. Pitowsky, Kluwer: University of Western Ontario 
Series in Philosophy of Science.\footnote{It is a pleasure to dedicate this paper 
to Jeff Bub, who has made such profound contributions to the philosophy of quantum 
theory. Though the paper is about classical, not quantum, mechanics, I  hope that 
with his love of geometry, he enjoys symplectic forms as much as inner products!}   

\end{center}

\begin{abstract}
This paper expounds the relations between continuous symmetries and conserved 
quantities, i.e. Noether's ``first theorem'', in both the Lagrangian and 
Hamiltonian frameworks for classical mechanics. This illustrates one of mechanics' 
grand themes: exploiting a  symmetry so as to reduce the number of variables 
needed to treat a problem.

I emphasise that, for both frameworks, the theorem is underpinned by the idea of 
cyclic coordinates; and that the Hamiltonian theorem is more powerful. The 
Lagrangian theorem's main ``ingredient'', apart from cyclic coordinates,   is the 
rectification of vector fields afforded by the local existence and uniqueness of 
solutions to ordinary differential equations. For the Hamiltonian theorem, the 
main extra ingredients are the asymmetry of the Poisson bracket, and the fact that 
a vector field generates canonical transformations iff it is Hamiltonian.   
\end{abstract}

\end{titlepage}

\tableofcontents

\newpage
\section{Introduction}\label{intro}
The strategy of simplifying a mechanical problem by exploiting a symmetry so as to 
reduce the number of variables is  one of classical mechanics' grand themes. It is 
theoretically deep, practically important, and recurrent in the history of the 
subject. Indeed, it occurs already in 1687, in Newton's solution  of the Kepler 
problem; (or more generally, the problem of two bodies exerting equal and opposite 
forces along the line between them). The symmetries are translations and 
rotations, and the corresponding conserved quantities are the linear and angular 
momenta. 

This paper will expound one central aspect of this large subject. Namely, the 
relations between continuous symmetries and conserved quantities---in effect, 
Noether's ``first theorem'': which I expound in both the Lagrangian and 
Hamiltonian frameworks, though confining myself to finite-dimensional systems. As 
we shall see, this topic is underpinned by the  theorems in elementary  Lagrangian 
and Hamiltonian mechanics about cyclic (ignorable) coordinates and their 
corresponding conserved momenta. (Again, there is a glorious history: these 
theorems were of course clear to these subjects' founders.) Broadly speaking, my 
discussion will make increasing use, as it proceeds, of the language of modern 
geometry. It will also emphasise Hamiltonian, rather than Lagrangian, mechanics: 
apart from mention of the Legendre transformation, the Lagrangian framework drops 
out wholly after Section \ref{NoetherLagGeomSubsub}.\footnote{It is worth noting 
the point, though I shall not exploit it,  that symplectic structure can be seen 
in the classical solution space of the Lagrangian framework; cf. (3) of Section 
\ref{geomlegetrsfmn}.}

There are several motivations for studying this topic. As regards physics, many of 
the ideas and results can be generalized to infinite-dimensional classical  
systems; and in either the original or the generalized form, they underpin 
developments in quantum theories. The topic also leads into  another important 
subject, the modern theory of symplectic reduction: (for a philosopher's 
introduction, cf. Butterfield (2006)).  As regards philosophy, the topic is a 
central focus for the  discussion of symmetry, which is both a long-established  
philosophical field and a currently active one: cf. Brading and Castellani (2003). 
(Some of the current interest relates to symplectic reduction, whose philosophical  
significance has been stressed recently, especially by Belot: Butterfield (2006) 
gives references.)

The plan of the paper is as follows. In Section \ref{Amsurvey}, I review the 
elements of the Lagrangian framework, emphasising the elementary theorem that 
cyclic coordinates yield conserved momenta, and introducing the modern geometric 
language in which mechanics is often cast. Then I review Noether's theorem in the 
Lagrangian framework (Section \ref{NoetherLag}). I emphasise how the theorem 
depends on two others:  the elementary theorem about cyclic coordinates, and  the 
local existence and uniqueness of solutions of ordinary differential equations. 
Then I introduce Hamiltonian mechanics, again emphasising how cyclic coordinates 
yield conserved momenta; and approaching canonical transformations through the 
symplectic  form (Section \ref{Hammechs}). This leads to Section \ref{PoissNoet}'s 
discussion of Poisson brackets; and thereby, of  the Hamiltonian version of 
Noether's theorem. In particular, we see what it would take to prove that this 
version is more powerful than (encompasses) the Lagrangian version. By the end of 
the Section, it only remains  to show that a vector field generates a 
one-parameter family of canonical transformations iff it is a Hamiltonian vector 
field. It turns out that we can show this without having to develop much of the 
theory of canonical transformations. We do so in the course of the final Section's   
account of the geometric structure  of Hamiltonian mechanics, especially the 
symplectic structure of a cotangent bundle (Section \ref{geomperspHam}). Finally, 
we end the paper by mentioning a generalized framework for  Hamiltonian mechanics 
which is crucial for symplectic reduction. This framework takes the Poisson 
bracket, rather than the symplectic form, as the basic notion; with the result 
that the state-space is, instead of a cotangent bundle, a generalization called a 
`Poisson manifold'.

\section{Lagrangian mechanics}\label{Amsurvey}

\subsection{Lagrange's equations}\label{Lageqsec}
We consider a mechanical system with $n$ configurational degrees of freedom (for 
short: {\em n freedoms}), described by the usual {\em Lagrange's equations}. These 
are $n$ second-order ordinary differential equations:
\be
\frac{d}{dt}(\frac{\partial L}{\partial \dot{q}^i}) - \frac{\partial L}{\partial 
q^i} = 0, \;\;\; i = 1,...,n ;
\label{eqn;lag}
\ee
where the  Lagrangian $L$ is the difference of the kinetic and potential energies: 
$L := K - V$. (We use $K$ for the kinetic energy, not the traditional $T$; for in 
differential geometry, we will use $T$ a lot, both for `tangent space' and 
`derivative map'.)
 
I should  emphasise at the outset that several  special assumptions are needed in 
order to deduce eq. \ref{eqn;lag} from Newton's second law, as applied to the 
system's component parts: (assumptions that tend to get forgotten in the geometric 
formulations that will dominate later Sections!) But I will not go into many 
details about this, since:\\
\indent (i): there is no single set of assumptions of mimimum logical strength 
(nor a single ``best package-deal'' combining  simplicity and mimimum logical 
strength);\\
\indent (ii): full discussions are available in many textbooks (or, from a 
philosophical viewpoint, in Butterfield 2004a: Section 3).\\
\indent I will just indicate a simple and commonly used sufficient set of 
assumptions. But owing to (i) and (ii), the details here will not be cited in 
later Sections.

Note first that if the system consists of $N$ point-particles (or bodies small 
enough to be treated as point-particles), so that a configuration is fixed by $3N$ 
cartesian coordinates, we may yet have $n < 3N$. For the system may be subject to 
constraints and we will require the $q^i$ to be independently variable. More 
specifically, 
let us assume that any constraints on the system are {\em holonomic}; i.e. each is 
expressible as an equation $f({r}^1,\dots,{r}^m) = 0$ among the coordinates 
${r}^k$ of the system's component parts; (here the $r^k$ could be the $3N$ 
cartesian coordinates of $N$ point-particles, in which case $m := 3N$). A set of 
$c$ such constraints can in principle be solved, defining a $(m - c)$-dimensional 
hypersurface $Q$ in the $m$-dimensional space of the $r$s; so that on the {\em 
configuration space} $Q$ we can define $n := m-c$ independent coordinates $q^i, i 
= 1,\dots,n$.

Let us also assume that any constraints on the system are: (i) {\em scleronomous}, 
i.e. independent of time, so that $Q$ is identified once and for all; (ii) {\em 
ideal}, i.e. the forces that maintain the constraints would do no work in any 
possible displacement consistent with the constraints and applied forces (a 
`virtual displacement'). Let us also assume that the forces  applied to the system 
are {\em monogenic}: i.e. the total work $\dd w$ done in an infinitesimal virtual 
displacement is integrable; its integral is the {\em work function} $U$.  (The 
term `monogenic' is due to Lanczos (1986, p. 30), but followed by others e.g. 
Goldstein et al. (2002, p. 34).) And let us assume that the system is {\em 
conservative}: i.e. the work function $U$ is independent of both the time and the 
generalized velocities ${\dot q}_i$, and depends only on the $q^i$: $U = 
U(q^1,\dots,q^n)$.

So to sum up: let us assume that the constraints are holonomic, scleronomous and 
ideal, and that the system is monogenic with a velocity-independent work-function. 
Now let us define $K$ to be the kinetic energy; i.e. in cartesian coordinates, 
with $k$ now labelling particles, $K := \Sigma_k \frac{1}{2}m_k{\bf v}^2_k$. Let 
us also define $V:= -U$ to be the potential energy, and set $L := K - V$. Then the 
above assumptions imply eq. \ref{eqn;lag}.\footnote{Though I shall not develop any 
details, there is of course a rich theory about these and related assumptions. One 
example, chosen with an eye to our later use of geometry, is that assuming 
scleronomous constraints, $K$ is readily shown to be a homogeneous quadratic form 
in the generalized velocities, i.e. of the form $K = \Sigma^n_{i,j} a_{ij}{\dot 
q}^i{\dot q}^j$; and so $K$ defines a metric on the configuration space.}

To solve mechanical problems, we need to integrate Lagrange's equations. Recall  
the idea from elementary calculus that $n$ second-order ordinary differential 
equations have a (locally) unique solution, once we are given $2n$ arbitrary 
constants.  Broadly speaking, this idea holds good for Lagrange's equations; and 
the $2n$ arbitrary constants can be given just as one would expect---as the 
initial configuration and generalized velocities $q^i(t_0), {\dot q}^i(t_0)$ at 
time $t_0$. More precisely: 
expanding the time derivatives in eq. \ref{eqn;lag}, we get
\be
\frac{\pl^2 L}{{\pl {\dot q}^j}{\pl {\dot q}^i}}{\ddot q}^j =
- \frac{\pl^2 L}{{\pl {q}^j}{\pl {\dot q}^i}}{\dot q}^j -
\frac{\pl^2 L}{{\pl t}{\pl {\dot q}^i}} + 
\frac{\pl L}{\pl {\dot q}^i}
\label{elpara=texpand}
\ee
so that the condition for being able to solve these equations to find the 
accelerations at some initial time $t_0$, ${\ddot q}^i(t_0)$, in terms of 
$q^i(t_0), {\dot q}^i(t_0)$ is that the {\em Hessian} matrix $\frac{\pl^2 L}{{\pl 
{\dot q}^i}{\pl {\dot q}^j}}$ be nonsingular. 
Writing the determinant as $\mid \;\; \mid$, and partial derivatives as 
subscripts, the condition is that:
\be
\mid \frac{\pl^2 L}{{\pl {\dot q}^j}{\pl {\dot q}^i}} \mid \;\;\; \equiv
\;\;\;
\mid L_{{\dot q}^j{\dot q}^i}\mid \;\;\; \neq \;\;\; 0 \;\; .
\label{nonzerohessian}
\ee
This {\em Hessian condition} holds in very many mechanical problems; and 
henceforth, we assume it. (If it fails, we enter the territory of constrained 
dynamics; for which cf. e.g. Henneaux and Teitelboim (1992, Chapters 1-5).) It 
underpins most of what follows: for it is  needed to define the Legendre 
transformation, by which we pass from Lagrangian to Hamiltonian mechanics.\\
\indent Of course, even with eq. \ref{nonzerohessian}, it is still in general hard 
{\em in practice} to solve for the ${\ddot q}^i(t_0)$: they are buried in the lhs 
of eq. \ref{elpara=texpand}. In (5) of Section \ref{tgtble}, this will  motivate 
the move to Hamiltonian mechanics.\footnote{This is not to say that  Hamiltonian 
mechanics makes all problems ``explicitly soluble'': if only! For a philosophical 
discussion of the various meanings of `explicit solution', cf. Butterfield (2004a: 
Section 2.1).} 

Given eq. \ref{nonzerohessian}, and so the accelerations at the initial time 
$t_0$, the basic theorem on the (local) existence and uniqueness of solutions of 
ordinary differential equations can be applied. (We will state this theorem in 
Section \ref{Noetsubsubsec} in connection with Noether's theorem.)

By way of indicating the rich theory that can be built from eq. \ref{eqn;lag} and 
\ref{nonzerohessian}, I mention one main aspect: the power of variational 
formulations. Eq. \ref{eqn;lag} are the Euler-Lagrange equations for the 
variational problem $\dd \int L \; dt = 0$; i.e. they are necessary and sufficient 
for the action integral $I = \int L \; dt$ to be stationary. But variational 
principles will play no further role in this paper; (Butterfield 2004 is a 
philosophical discussion).

But our main concern, here and throughout this paper, is how symmetries yield 
conserved quantities, and thereby reduce the number of variables that need to be 
considered in solving a problem. In fact, we are already in a position to prove 
Noether's theorem, to the effect that  any (continuous) symmetry of the Lagrangian 
$L$ yields a conserved quantity. But we postpone this to Section \ref{NoetherLag}, 
until we have developed some more notions, especially geometric ones. 

We begin with the idea of generalized momenta, and the result that the generalized  
momentum of any cyclic coordinate is a constant of the motion: though very simple, 
this result is the basis of  Noether's theorem. Elementary examples prompt the 
definition of the {\em generalized}, or {\em canonical}, {\em momentum}, $p_i$,  
{\em conjugate} to a coordinate $q^i$ as: $\frac{\partial L}{\partial {\dot 
q^i}}$; (this was first done by Poisson in 1809). Note that $p_i$ need not have 
the dimensions of momentum: it will not if $q^i$ does not have the dimension 
length. 
So Lagrange's equations can be written:
\be
\frac{d}{dt} p_i = \frac{\pl L}{\pl {q^i}} \;\; ;
\label{LagWithpj}
\ee
We say a coordinate $q^i$ is {\em cyclic} if $L$ does not depend on $q^i$. (The 
term comes from the example of an angular coordinate of a particle subject to a 
central force. Another term is: {\em ignorable}.) Then the Lagrange equation for a 
cyclic coordinate, $q^n$ say,  becomes ${\dot p}_n = 0$, implying
\be
p_n = \mbox {constant, $c_n$ say}.
\label{pnconstant}
\ee
So: the generalized momentum conjugate to a cyclic coordinate is a constant of the 
motion.

It is straightforward to show that this simple result encompasses the elementary 
theorems of the conservation of momentum, angular momentum and energy: this last 
corresponding to time's being a cyclic coordinate. As a simple example, consider 
the angular momentum of a free particle. The Lagrangian is, in spherical polar 
coordinates,
\be
L = \frac{1}{2}m({\dot r}^2 + r^2{\dot \theta}^2 + r^2{\dot \phi}^2\sin^2\theta) 
\label{LagSpherl}
\ee 
so that ${\pl L}/{\pl \phi} = 0$. So the conjugate momentum
\be
\frac{\pl L}{\pl {\dot \phi}} = mr^2{\dot \phi}\sin^2\theta \; ,
\label{AMfreeSpherl}
\ee
which is the angular momentum about the $z$-axis, is conserved.

\subsection{Geometrical perspective}\label{geomperspLag}

\subsubsection{Some restrictions of scope}\label{limitsgeomperspLag}
I turn to give a brief description of the elements of Lagrangian mechanics in 
terms of modern differential geometry. Here `brief' indicates that:\\
\indent (i): I will assume without explanation various geometric notions, in 
particular: manifold, vector, 1-form (covector), metric, Lie derivative and 
tangent bundle.\\
\indent (ii): I will disregard issues about degrees of smoothness: all manifolds, 
scalars, vectors etc. will be assumed to be as smooth as needed for the context.\\
\indent  (iii): I will also simplify by speaking ``globally, not locally''. I will 
speak as if the  scalars, vector fields etc. are defined on a whole manifold; when 
in fact all that we can claim in application to most systems is a corresponding 
local statement---because for example, differential equations are guaranteed  the 
existence and uniqueness only of a {\em local} solution.\footnote{A note for {\em 
afficionados}. Of the three main pillars of elementary differential geometry---the 
implicit function theorem, the local existence and uniqueness of solutions of 
ordinary differential equations, and Frobenius' theorem---this paper will use the 
first only implicitly (!), and the second explicitly in Sections \ref{NoetherLag} 
and \ref{Hammechs}. The third will not be used.}

We begin by assuming that the configuration space (i.e. the constraint surface) 
$Q$ is a manifold. The physical state of the system, taken as a pair of 
configuration and generalized velocities, is represented by a point in the tangent 
bundle $TQ$ (also known as `velocity phase space'). That is, writing $T_x$ for the 
tangent space at $x \in Q$, $TQ$ has points $(x, \tau), x \in Q, \tau \in T_x$. We 
will of course often work with the natural coordinate systems on $TQ$ induced by  
coordinate systems $q$ on $Q$; i.e. with the $2n$ coordinates $(q,{\dot q}) \equiv 
(q^i, {\dot q}^i)$.   

The main idea of the geometric perspective is  that this tangent bundle is the 
arena for Lagrangian mechanics. So various previous notions and results are now 
expressed in terms of the tangent bundle. In particular, the Lagrangian is a 
scalar function $L:TQ \rightarrow \mathR$ which ``determines everything''.  And 
the conservation of the  generalized momentum $p_n$ conjugate to a cyclic 
coordinate $q_n$, $p_n \equiv p_n(q, {\dot q}) = c_n$, means that the motion of 
the system is confined to a level set $p_n^{-1}(c_n)$: where this level set is a 
$(2n-1)$-dimensional sub-manifold of $TQ$.

But I must admit at the outset that working with $TQ$ involves limiting our 
discussion to (a) time-independent Lagrangians and (b) time-independent coordinate 
transformations.\\
\indent (a): Recall Section \ref{Lageqsec}'s assumptions that secured eq. 
\ref{eqn;lag}. Velocity-dependent potentials and-or  rheonomous constraints would  
prompt one to use what is often called the  `extended configuration space' $Q 
\times \mathR$, and-or the `extended velocity phase space' $TQ \times \mathR$.\\
\indent  (b): So would  time-dependent coordinate  transformations. This is  a 
considerable limitation from a philosophical viewpoint, since it excludes boosts, 
which are central to the philosophical discussion of  spacetime symmetry groups, 
and especially of  relativity principles. To give the simplest example: the 
Lagrangian of a free particle is just its kinetic energy, which can be made zero 
by transforming to the particle's rest frame; i.e.  it is not invariant under 
boosts. 

\subsubsection{The tangent bundle}\label{tgtble}
With these limitations admitted, we now describe Lagrangian mechanics on $TQ$, in 
five extended comments.

  (1): {\em  $2n$ first-order equations; the Hessian again}:---\\
The Lagrangian equations  of motion are now $2n$ {\em first-order} equations for 
the functions $q^i(t), {\dot q}^i(t)$, falling in to two groups:\\
\indent \indent (a) the $n$ equations eq. \ref{elpara=texpand}, with the ${\ddot 
q}^i$ taken as the time derivatives of ${\dot q}^i$ with respect to $t$; i.e. we 
envisage using the Hessian condition eq. \ref{nonzerohessian} to solve eq. 
\ref{elpara=texpand} for the ${\ddot q}^i$, hard though this usually is to do in 
practice;\\
\indent \indent (b) the $n$ equations ${\dot q}^i = \frac{d q^i}{dt}$.

 (2): {\em Vector fields and solutions}:---\\
\indent \indent (a): These $2n$ first-order equations are equivalent to a vector 
field on $TQ$: the `dynamical vector field', or for short the `dynamics'. I write 
it as $D$ (to distinguish it from the generic vector field $X,Y,...$).\\  
\indent\indent (b): In the natural coordinates $ (q^i, {\dot q}^i)$, the vector 
field $D$ is expressed as
\be
D = {\dot q}^i \frac{\pl}{\pl q^i} + {\ddot q}^i \frac{\pl}{\pl {\dot q}^i} \;\; ;
\label{LagDelta}
\ee
and the rate of change of any dynamical variable $f$, taken as a scalar function 
on $TQ$, $f(q,{\dot q}) \in \mathR$ is given by
\be
 \frac{df}{dt} = {\dot q}^i \frac{\pl f}{\pl q^i} + {\ddot q}^i \frac{\pl f}{\pl 
{\dot q}^i} = D(f) .
\label{LagDeltaforf}
\ee   
\indent\indent (c): So the Lagrangian $L$  determines the dynamical vector field 
$D$, and so (for given initial $q,{\dot q}$) a (locally unique) solution: an 
integral curve of $D$, $2n$ functions of time $q(t), {\dot q}(t)$ (with the first 
$n$ functions determining the latter). This separation of solutions/trajectories 
within $TQ$ is important for the visual and qualitative understanding of 
solutions. 

 (3): {\em Canonical momenta are 1-forms}:---\\
Any point transformation, or any coordinate transformation $(q^i) \rightarrow 
(q'^i)$, in the configuration manifold $Q$, induces a basis-change in the tangent 
space $T_q$ at $q \in Q$. Consider any vector $\tau \in T_q$ with components 
${\dot q}^i$ in coordinate system $(q^i)$ on $Q$, i.e. $\tau = \frac{d}{dt} = 
{\dot q}^i \frac{\pl}{\pl q^i}$; (think of a motion through configuration $q$ with 
generalized velocity $\tau$). Its  components ${\dot q}^{'i}$ in the coordinate 
system $(q'^i)$ (i.e. $\tau = {\dot q}^{'i} \frac{\pl}{\pl q^{'i}}$) are given by 
applying the chain rule to $q^{'i} = q^{'i}(q^k)$:
\be
{\dot q}^{'i} \equiv \frac{\pl q'^i}{\pl q^k}{\dot q}^k .
\label{chainforpttrsfmn}
\ee
so that we can ``drop the dots'': 
\be
\frac{\pl {\dot q}^{'i}}{\pl {\dot q}^j} = \frac{\pl {q}^{'i}}{\pl {q}^j} .
\label{dropthedots}
\ee
One easily checks, using eq. \ref{dropthedots}, that for any  $L$, the canonical 
momenta $p_i := \frac{\pl L}{\pl {\dot q}^i}$ form a 1-form on $Q$, transforming 
under $(q^i) \rightarrow (q'^i)$ by:
\be
p'_i := \frac{\pl L'}{\pl {\dot q}^{'i}} = \frac{\pl q^k}{\pl { q}^{'i}} \frac{\pl 
L}{\pl {\dot q}^{k}} \equiv \frac{\pl q^k}{\pl { q}^{'i}} p_k
\label{psareoneform}
\ee
That is, the canonical momenta defined by $L$ form a 1-form field on $Q$. (We will 
later describe this as  a cross-section of the cotangent bundle.)

 (4): {\em Geometric formulation of Lagrange's equations}:---\\
We can formulate Lagrange's equations in a coordinate-independent way, by using 
three ingredients, namely:\\
\indent (i): $L$ itself (a scalar, so coordinate-independent);\\
\indent (ii):  the vector field $D$ that $L$ defines; and\\
\indent (iii): the 1-form on $TQ$ defined locally, in terms of the natural 
coordinates $(q^i,{\dot q}^i)$, by
\be
\theta_L := \frac{\pl L}{\pl {\dot q}^i}dq^i \;\;.
\label{defpdq}
\ee
(So the coefficients of $\theta_L$ for the other $n$ elements of the dual basis, 
the $d{\dot q}^i$ are defined to be zero.) This 1-form is called the {\em 
canonical 1-form}. We shall see that it plays a role in Noether's theorem, and is 
centre-stage in  Hamiltonian mechanics.\\
\indent We combine these three ingredients using the idea of the Lie derivative of 
a 1-form along a vector field.

We will write the  Lie derivative of $\theta_L$ along the vector field  $D$ on 
$TQ$,  as ${\cal L}_{D} \theta_L$. (It is sometimes written as $L$; but we need 
the symbol $L$ for the Lagrangian---and later on, for left translation.) By the 
Leibniz rule, ${\cal L}_{D} \theta_L$ is:
\be
{\cal L}_{D} \theta_L =   ({\cal L}_{D} \frac{\pl L}{\pl {\dot q}^i} ) dq^i +  
\frac{\pl L}{\pl {\dot q}^i}{\cal L}_{D}(dq^i) \;\;.
\label{LieOfpdq}
\ee
But the Lie derivative of any scalar function $f:TQ \rightarrow \mathR$ along any 
vector field $X$ is just $X(f)$; and for the dynamical vector field $D$, this is 
just ${\dot f} = \frac{\pl f}{\pl q^i}{\dot q}^i + \frac{\pl f}{\pl {\dot 
q}^i}{\ddot q}^i$. So we have
\be
{\cal L}_{D} \theta_L =   (\frac{d}{dt} \frac{\pl L}{\pl {\dot q}^i} ) d q^i +  
\frac{\pl L}{\pl {\dot q}^i} d {\dot q}^i  \;\; .
\label{LieOfpdqdbydt}
\ee
Rewriting the first term by the Lagrange equations, we get
\be
{\cal L}_{D} \theta_L =   (\frac{\pl L}{\pl {q}^i} ) dq^i +  \frac{\pl L}{\pl 
{\dot q}^i}d{\dot q}^i \equiv dL \;\; .
\label{coordindpdtLag}
\ee
We can conversely deduce the familiar Lagrange equations from eq. 
\ref{coordindpdtLag}, by taking coordinates. So we conclude that these equations'  
coordinate-independent form is:\\
\be
{\cal L}_{D} \theta_L =  dL \;\; .
\label{coordindpdtLag2}
\ee

 (5): {\em Towards the Hamiltonian framework}:---\\
 Finally, a comment about the  Lagrangian framework's limitations as regards 
solving problems, and how they prompt the transition to Hamiltonian mechanics.\\
\indent Recall the remark at the end of Section \ref{Lageqsec}, that the $n$ 
equations  eq. \ref{elpara=texpand} are in general hard to solve for the ${\ddot 
q}^i(t_0)$: they lie buried in the left hand side  of eq. \ref{elpara=texpand}. On 
the other hand, the $n$ equations ${\dot q}^i = \frac{d q^i}{dt}$ (the second 
group of $n$ equations  in (1) above) are as simple as can be. \\
\indent This makes it natural to seek another $2n$-dimensional  space of 
variables, $\xi^{\al}$ say ($\al=1,...,2n$), in which:\\
\indent \indent (i): a motion is described by first-order equations, so that we 
have the same advantage as in $TQ$ that a unique trajectory passes through each 
point of the space; but in which\\
\indent \indent  (ii): all $2n$ equations have the simple form $\frac{d 
\xi^{\al}}{dt} = f_{\al}(\xi^1,...\xi^{2n})$ for some set of functions $f_{\al} 
(\al=1,...,2n)$.\\
\indent \indent Indeed, Hamiltonian mechanics provides exactly such a space: it is 
usually  the cotangent bundle of the configuration manifold, instead of its 
tangent bundle. But before turning to that, we expound Noether's theorem in the 
current Lagrangian framework.

\section{Noether's theorem in Lagrangian mechanics}\label{NoetherLag}
\subsection{Preamble: a modest plan}\label{NoetherPreamble}
Any discussion of symmetry in Lagrangian mechanics must include a treatment of 
``Noether's theorem''. The scare quotes are to indicate that there is more than 
one Noether's theorem. Quite apart from Noether's work in other branches of 
mathematics, her paper (1918) on symmetries and conserved quantities in Lagrangian 
theories has several theorems. I will be concerned {\em only} with applying her 
first theorem to finite-dimensional systems. In short: it provides, for any 
continuous symmetry of a system's Lagrangian, a conserved quantity called the 
`momentum  conjugate to the symmetry'.

\indent I stress at the outset that the great majority of subsequent applications 
and commentaries (also for her other theorems, besides her first) are concerned 
with versions of the theorems for infinite (i.e. continuous) systems. In fact, the 
context of Noether's investigation was contemporary debate about how to understand 
conservation principles and symmetries in the ``ultimate classical continuous 
system'', viz. gravitating matter as described by Einstein's general relativity. 
This theory can be given a Lagrangian formulation: that is, the equations of 
motion, i.e. Einstein's field equations, can be deduced from a Hamilton's 
Principle with an appropriate Lagrangian.  The contemporary debate was especially 
about the conservation of energy and the principle of general covariance (also 
known as: diffeomorphism invariance). General covariance prompts one to consider 
how a variational principle transforms under  spacetime coordinate transformations 
that are arbitrary, in the sense of varying from point to point. This leads to the 
idea of ``local'' symmetries, which since Noether's time has been immensely 
fruitful in both classical and quantum physics, and in both a Lagrangian and 
Hamiltonian framework.\footnote{Cf. Brading and Castellani (2003). Apart from 
papers specifically about Noether's theorem, this anthology's papers by Wallace, 
Belot and Earman (all 2003) are closest to this paper's concerns.}

\indent So I agree that from the perspective of Noether's work, and its enormous 
later development, this Section's application of the first theorem to 
finite-dimensional systems is, as they say, ``trivial''. Furthermore, this 
application is easily understood, {\em without} having to adopt that perspective, 
or even having to consider infinite systems.  In other words:  its statement and 
proof are natural, and simple, enough that the nineteenth century masters of 
mechanics, like Hamilton, Jacobi and Poincar\'{e}, would certainly recognize it in 
their own work---allowing of course for adjustments to modern language. In fact, 
versions of it for the Galilei group of Newtonian mechanics and the Lorentz group 
of special relativity were published a few years before Noether's paper; (Brading 
and Brown (2003, p. 90); for details, cf. Kastrup (1987)).\footnote{Here again, 
`versions of it' needs scare-quotes. For in what follows, I shall be more limited 
than these proofs, in two ways. (1): I limit myself, as I did in Section 
\ref{limitsgeomperspLag}, both to time-independent Lagrangians and to 
time-independent transformations: so my discussion does not encompass boosts. (2): 
I will take a symmetry of $L$ to require that $L$ be the {\em very same}; whereas 
some treatments allow the addition to $L$ of the time-derivative of a function 
$G(q)$ of the coordinates $q$---since such a time-derivative  makes no   
difference to the Lagrange  equations.}

\indent Nevertheless, it is worth expounding the finite-system version of 
Noether's first theorem. For:\\
\indent (i): It  generalizes Section \ref{Lageqsec}'s result about cyclic 
coordinates, and thereby the elementary theorems of the conservation of momentum, 
angular momentum and energy which that result encompasses. The main generalization 
is that the theorem does not assume we have identified a cyclic coordinate. But on 
the other hand: every symmetry in the Noether  sense will arise from a cyclic 
coordinate in some system $q$ of generalized coordinates. (As we will see,  this 
follows from the local existence and uniqueness of solutions of ordinary 
differential equations.) \\
\indent (ii): This exposition will also prepare the way for our discussion of 
symmetry and conserved quantities in Hamiltonian mechanics.\footnote{Other 
expositions of Noether's theorem for finite-dimensional Lagrangian mechanics 
include: Arnold (1989: 88-89), Desloge (1982: 581-586), Lanczos (1986: 401-405: 
emphasizing the variational perspective) and Johns (2005: Chapter 13). Butterfield 
(2004a, Section 4.7) is a  more detailed version of this Section. Beware: though 
many textbooks of Hamiltonian mechanics cover the Hamiltonian version of Noether's 
theorem (which, as we will see, is stronger), they often do not label it as such; 
and if they do label it, they often do not relate it clearly to the Lagrangian 
version.}

In this exposition, I will also discuss {\em en passant} the distinction 
between:\\
\indent (i) the notion of symmetry at work in Noether's theorem, i.e. a symmetry 
of $L$, often called a {\em variational symmetry}; and\\
\indent  (ii) the notion of a symmetry of the set of solutions of a differential 
equation: often called a {\em dynamical symmetry}. This notion applies to all 
sorts of differential equations, and systems of them; not just to those with the 
form of Lagrange's  equations (i.e. derivable from an variational principle). In 
short, this sort of symmetry is a map that sends any solution of the given 
equation(s)  (in effect: a dynamically possible history of the system---a curve in 
the state-space) to some other solution. Finding such symmetries, and groups of 
them, is a central part of the modern theory of integration of differential 
equations (both ordinary and partial).\\
\indent Broadly speaking, this notion is more general than that of a symmetry of 
$L$. Not only does it apply to many other sorts of differential equation. Also, 
for Lagrange's equations: a symmetry of $L$ is (with one {\em caveat}) a symmetry 
of the solutions, i.e. a dynamical symmetry---but the converse is 
false.\footnote{An excellent account of this modern integration theory, covering 
both ordinary and partial differential equations, is given by Olver (2000). He 
also covers the Lagrangian case (Chapter 5 onwards), and gives many historical 
details especially about Lie's pioneering contributions.}  
  
\indent In this Section, the plan is as follows. We define:\\
\indent (i): a {\em (continuous) symmetry} as a vector field (on the configuration 
manifold $Q$) that generates a family of transformations  under which the 
Lagrangian is invariant;  (Section \ref{VecfieldsSymmies});\\
\indent (ii): the {\em momentum conjugate to a vector field}, as (roughly) the 
rate of change of the Lagrangian with respect to the ${\dot q}$s in the direction 
of the vector field; (Section \ref{NoetherConjMomm}).\\ 
\indent These two definitions lead directly to Noether's theorem (Section 
\ref{Noetsubsubsec}): after all the stage-setting, the proof  will be a one-liner 
application of Lagrange's equations.

\subsection{Vector fields and  symmetries---variational and 
dynamical}\label{VecfieldsSymmies}
I need to expound three topics:\\
\indent (1): the idea of a vector field on the configuration manifold $Q$; and how 
to lift it to $TQ$;\\
\indent (2): the definition of a variational symmetry;  \\
\indent (3): the contrast between (2) and the idea of dynamical symmetry.\\
Note that, as in previous Sections, I will often speak, for  simplicity, 
``globally, not locally'', i.e. as if the relevant scalar functions, vector fields 
etc. are defined on all of $Q$ or $TQ$. Of course, they need not be.

\subsubsection{Vector fields on $TQ$; lifting fields from $Q$ to 
$TQ$}\label{vfsonTQ}  
We recall first that a differentiable vector field on $Q$ is represented in a 
coordinate system $q = (q^1,\dots,q^n)$ by $n$ first-order ordinary differential 
equations
\be
\frac{d q^i}{d \epsilon} = f^i(q^1,\dots,q^n) \;\; .
\label{vecfieldintuitive}
\ee
A vector field generates a one-parameter family of active transformations: viz. 
passage along the vector field's integral  curves, by a varying 
parameter-difference $\epsilon$. The vector field is called the {\em infinitesimal 
generator} of the family. It is common to write the parameter as $\tau$, but in 
this Section we use $\epsilon$ to avoid confusion with  $t$, which often 
represents the time.

Similarly, a vector field defined on 
$TQ$ corresponds to a system of $2n$ ordinary differential equations, and 
generates an active transformation of $TQ$.
But I will consider only vector fields on $TQ$ that mesh with the structure of 
$TQ$ as a tangent bundle, in the sense that they are induced by vector fields on 
$Q$, in the following natural way.

This induction has two ingredient ideas.\\
\indent  First, any curve in $Q$ (representing a possible state of motion) defines 
a corresponding curve in $TQ$, because the functions $q^i(t)$ define the functions 
${\dot q}^i(t)$. (Here $t$ is the parameter of the curve.) More formally: given 
any curve in configuration space, $\phi: I \subset \mathR \rightarrow Q$, with 
coordinate expression in the $q$-system $t \in I \mapsto q(\phi(t)) \equiv q(t) = 
q^i(t)$, we define its {\em extension} to $TQ$ to be the curve $\Phi: I \subset 
\mathR \rightarrow TQ$ given in the corresponding coordinates by $q^i(t),{\dot 
q}^i(t)$.\\
\indent Second, any vector field $X$ on $Q$ generates displacements in any 
possible state of motion, represented by a curve in $Q$ with coordinate expression 
$q^i = q^i(t)$. Namely: for a given value of the parameter  $\epsilon$, the 
displaced state of motion is represented by the curve in Q
\be
q^i(t) + \epsilon X^i(q^i(t)) \;\; .
\label{displacedinQ}
\ee 
\indent Putting these ingredients together: we first displace a curve within $Q$, 
and then extend the result to $TQ$. Namely, the extension to $TQ$ of the (curve 
representing) the displaced state of motion is  given by the $2n$ functions, in 
two groups each of $n$ functions, for the $(q,{\dot q})$ coordinate system 
\be
q^i(t) + \epsilon X^i(q^i(t)) \;\; \mbox{ and } \;\;
{\dot q}^i(t) + \epsilon Y^i(q^i(t), {\dot q}^i) \;\; ;
\ee
where $Y$ is defined to be the vector field on $TQ$ that is the derivative along 
the original state of motion of $X$. That is: 
\be
Y^i(q,{\dot q}) := \frac{d X^i}{dt} = \Sigma_j \; \frac{\pl X^i}{\pl q^j}{\dot 
q}^j.
\label{defineY}
\ee
Thus displacements by a vector field within $Q$  are lifted to $TQ$. The vector 
field $X$ on $Q$ lifts to $TQ$ as $(X,\frac{d X}{dt})$; i.e. it lifts to the 
vector field that sends a point $(q^i,{\dot q}^i) \in TQ$ to $(q^i + \epsilon X^i, 
{\dot q}^i + \epsilon \frac{d X^i}{dt})$.\footnote{I have discussed this in terms 
of some system $(q,{\dot q})$ of coordinates. But the definitions of extensions 
and displacements are in fact coordinate-independent. 
 Besides, one can show that
the operations of displacing a curve within $Q$, and extending it to $TQ$, commute 
to first order in $\epsilon$: the result is the same for either order of the 
operations.}
 
\subsubsection{The definition of variational  symmetry}\label{defvarnlsym}
To define variational symmetry, I begin with the integral  notion and then give 
the differential notion. The idea is that the Lagrangian $L$, a scalar $L:TQ 
\rightarrow \mathR$, should be invariant under all the  elements of a 
one-parameter family of active transformations $\theta_{\epsilon} : \epsilon \in I 
\subset \mathR$: at least in a neighbourhood of the identity map corresponding  to 
$\epsilon = 0$, $\theta_0 \equiv id_U$. (Here $U$ is some open subset of $TQ$, 
maybe not all of it.)\\
\indent That is, we define the family $\theta_{\epsilon} : \epsilon \in I \subset 
\mathR$ to  be a {\em variational  symmetry} of $L$ if $L$ is invariant under the 
transformations: $L = L \circ \theta_{\epsilon}$, at least around $\epsilon = 0$.  
(We could use the correspondence between active and passive transformations to 
recast this definition, and what follows, in terms of a passive notion of symmetry 
as sameness of $L$'s functional form in different coordinate systems. I leave this 
as an exercise! Or cf. Butterfield (2004a: Section 4.7.2).)

For the differential notion of variational symmetry, we of course use the idea of 
a vector field. But we also impose Section \ref{vfsonTQ}'s restriction to vector 
fields on $TQ$ that are induced by vector fields on $Q$. So we define a vector 
field $X$ on $Q$ that generates a family of active transformations 
$\theta_{\epsilon}$ on $TQ$ to be a variational symmetry of $L$ if the first 
derivative of $L$ with respect to $\epsilon$ is zero, at least around $\epsilon = 
0$.  More precisely: writing  
\be
L \circ \theta_{\epsilon} = L(q^i + \epsilon X^i, {\dot q}^i + \epsilon Y^i) 
\;\; 
\mbox{ with } \; Y^i = \Sigma_j \;\frac{\pl X^i}{\pl q^j}{\dot q^j} \; ,
\label{definevarnlsymmy}
\ee    
we say $X$ is a {\em variational symmetry} iff the first derivative of $L$ with 
respect to $\epsilon$ is zero (at least around $\epsilon = 0$). That is:  $X$ is a 
variational symmetry iff
\be
\Sigma_i \; X^i \frac{\pl L}{\pl q^i} \; + \; \Sigma_i \; Y^i \frac{\pl L}{\pl 
{\dot q^i}} \;\; = 0 \;\; 
\mbox{ with } \; Y^i = \Sigma_j \;\frac{\pl X^i}{\pl q^j}{\dot q^j} \; .
\label{definevarnlsymy2}
\ee

\subsubsection{A contrast with dynamical symmetries}\label{CfDynlsym}
The general notion of a dynamical symmetry, i.e. a symmetry of some equations of 
motion (whether Euler-Lagrange or not), is not needed for Section 
\ref{Noetsubsubsec}'s presentation of Noether's theorem. But the notion is so 
important that I must mention it,  though  only to contrast it with variational 
symmetries.

The general definition is roughly as follows. Given any system of differential 
equations, ${\cal E}$ say, a {\em dynamical symmetry} of the system is an active 
transformation $\zeta$ on the system $\cal E$'s space of both independent 
variables, $x_j$ say, and dependent variables $y^i$ say, such that any solution of 
$\cal E$, $y^i = f^i(x_j)$ say, is carried to another solution. For a precise 
definition, cf. Olver (2000: Def. 2.23, p. 93), and his ensuing discussion of the 
induced action (called `prolongation') of the transformation $\zeta$ on the spaces 
of (in general, partial) derivatives of the $y$'s with respect to the $x$s (i.e. 
jet spaces).\\
\indent As I said in Section \ref{NoetherPreamble}, groups of symmetries in this 
sense play a central role in the modern theory of differential equations: not just 
in finding new solutions, once given a solution, but also in integrating the 
equations. For some main theorems stating criteria (in terms of prolongations) for 
groups of symmetries, cf. Olver (2000: Theorem 2.27, p. 100, Theorem 2.36, p. 110, 
Theorem 2.71, p. 161).

But for present purposes, it is enough to state the rough idea of a one-parameter 
group of dynamical symmetries (without details about prolongations!) for  
Lagrange's equations in the familiar form,  eq. \ref{eqn;lag}.

 In this simple case, there is just one independent variable $x := t$, so that:\\
\indent (a): we are considering ordinary, not partial, differential equations, 
with $n$ dependent variables $y^i := q^i(t)$.\\
\indent (b): prolongations correspond to lifts of maps on $Q$ to maps on $TQ$; cf. 
Section \ref{vfsonTQ}.\\
\indent Furthermore, in line with the discussion following Lagrange's equations 
eq. \ref{eqn;lag}, the time-independence of the Lagrangian (time being a cyclic 
coordinate) means we can define dynamical symmetries $\zeta$ in terms of active 
transformations  on the tangent bundle, $\theta: TQ \rightarrow TQ$, that are 
lifted from active transformations on $Q$.  In effect, we define such a map 
$\zeta$ by just adjoining to any such $\theta: TQ \rightarrow TQ$ the identity map 
on the time variable $id: t \in \mathR \mapsto t$. (More formally: $\zeta: (q, 
{\dot q}, t) \in TQ \times \mathR \mapsto (\theta(q, {\dot q}), t) \in TQ \times 
\mathR$.)\\
\indent Then we define in the usual way what it is for a one-parameter family of 
such maps   $\zeta_s : s \in I \subset \mathR$ to be a (local) one-parameter group 
of dynamical symmetries (for Lagrange's equations eq. \ref{eqn;lag}): namely, if 
any solution curve $q(t)$ (equivalently: its extension $q(t),{\dot q}(t)$ to $TQ$) 
of the Lagrange equations is carried by each $\zeta_s$ to another solution curve, 
with the $\zeta_s$ for different $s$ composing in the obvious way, for $s$ close 
enough to $0 \in I$.\\
\indent And finally: we also define (in a manner corresponding to the discussion 
at the end of Section \ref{defvarnlsym}) a differential, as against integral, 
notion of dynamical symmetry. Namely, we say a vector field $X$ on $Q$ is a 
dynamical symmetry if its lift to $TQ$ (more precisely: its lift, with the 
identity map on the time variable adjoined) is the infinitesimal generator of such 
a one-parameter family $\zeta_s$.

For us, the important point is that this notion of a dynamical symmetry is {\em 
different} from Section \ref{defvarnlsym}'s notion of a variational 
symmetry.\footnote{Since the Lagrangian $L$ is especially associated with 
variational principles, while the dynamics is given by equations of motion, 
calling  Section \ref{defvarnlsym}'s notion `variational symmetry', and this 
notion `dynamical symmetry' is a good and widespread usage. But beware: it is not 
universal.} As I announced in Section \ref{NoetherPreamble}, a variational 
symmetry is (with one {\em caveat}) a dynamical symmetry---but the converse is 
false. Fortunately, the same simple example will serve both to show the subtlety 
about the first implication, and as a counterexample to the converse implication. 
This example is the two-dimensional harmonic oscillator.\footnote{All the material 
to the end of this Subsection is drawn from Brown and Holland (2004a); cf. also 
their (2004). The present use of the harmonic oscillator example also occurs in 
Morandi et al (1990: 203-204).}

The usual Lagrangian is, with cartesian coordinates written as $q$s, and the 
contravariant indices written for clarity as subscripts:
\be
L_1 =  \frac{1}{2}\left[{\dot q_1}^2 + {\dot q_2}^2 \; - \; \omega^2 (q^2_1 + 
q^2_2) \right] \;\;   ;
\label{Lag2DHO;eqns}
\ee
giving as Lagrange equations:
\be
{\ddot q}_i + \omega^2 q_i = 0 \;\; ,  \;\; i = 1,2 .
\label{eqnsfor1freq2DHO}
\ee  
But these Lagrange equations, i.e. the same dynamics, are also given by
\be
L_2 = {\dot q_1}{\dot q_2} - \omega^2 q_1 q_2 \;\; .
\label{OddLag2DHO}
\ee
The rotations in the plane are of course a variational symmetry  of $L_1$, and a 
dynamical symmetry of eq. \ref{eqnsfor1freq2DHO}. But they are {\em not} a 
variational symmetry of $L_2$. So a dynamical symmetry need not be a variational 
one. Besides, these equations contain another example to the same effect. Namely, 
the ``squeeze'' transformations 
\be
q'_1 := e^{\eta}q_1 \; , \; q'_2 := e^{-\eta}q_2
\label{squeeze}
\ee
are a dynamical symmetry of eq. \ref{eqnsfor1freq2DHO}, but not a variational 
symmetry of $L_1$. So again: a dynamical symmetry need not be a variational 
one.\footnote{In the light of this, you might ask about a more restricted 
implication: viz. must every dynamical symmetry of a set of equations of motion be 
a variational symmetry of {\em some or other Lagrangian} that yields the given 
equations as the Euler-Lagrange equations of Hamilton's Principle? Again, the 
answer is No for the simple reason that there are many (sets of) equations of 
motion  that are not Euler-Lagrange equations of {\em any} Lagrangian, and yet 
have dynamical symmetries.

Wigner (1954) gives an example. The general question of under what conditions is a 
set of ordinary differential equations the Euler-Lagrange equations of some 
Hamilton's Principle is the {\em inverse problem} of Lagrangian mechanics. It is a 
large subject with a long history; cf. e.g. Santilli (1979), Lopuszanski (1999).}  

I turn to the first implication: that every variational symmetry is a dynamical 
symmetry. This is true: general and abstract proofs (applying also to continuous 
systems i.e. field theories) can be found in Olver (2000: theorem 4.14, p. 255; 
theorem 4.34, p. 278; theorem 5.53, p. 332).\\
\indent  But beware of a condition of the theorem. (This is the {\em caveat} 
mentioned at the end of Section \ref{NoetherPreamble}.) The theorem requires that 
all the variables $q$ (for continuous systems: all the fields $\phi$) be subject 
to Hamilton's Principle. The need for this condition  is shown by rotations in the 
plane, which are a variational symmetry  of the  familiar Lagrangian  $L_1$ above. 
But it is easy to show that such a rotation is a dynamical symmetry of one of the 
Lagrange equations, say the equation for the variable $q_1$ 
\be
{\ddot q}_1 + \omega^2 q_1 = 0 \;\;, 
\label{eqnsforq1for1freq2DHO}
\ee
{\em only if} the corresponding Lagrange equation holds for $q_2$.

\subsection{The conjugate momentum of a vector field}\label{NoetherConjMomm}
Now we define {\em the momentum conjugate to a vector field} $X$ to be the scalar 
function on $TQ$:
\be
p_X: TQ \rightarrow \mathR \;\; ; \;\; p_X = \Sigma_i \; X^i \frac{\pl L}{\pl 
{\dot q}^i}
\label{defconjugmommX}
\ee
(For a time-dependent Lagrangian, $p_X$ would be a scalar function on $TQ \times 
\mathR$, with $\mathR$ representing time.)

\indent We shall see in the next Subsection's examples that this definition 
generalizes in an appropriate way Section \ref{Lageqsec}'s definition of the 
momentum conjugate to a coordinate $q$.\\
\indent But first note that it is an {\em improvement} in the sense that, while 
the momentum conjugate to a coordinate $q$ depends on the choice made for the 
other coordinates, the momentum $p_X$ conjugate to a vector field $X$ is 
independent of the coordinates chosen. Though this point is not needed in order to 
prove Noether's theorem, here is the proof.

We first apply the chain-rule to $L = L(q'(q), {\dot q}'(q,{\dot q}))$ and eq. 
\ref{dropthedots}  (``cancellation of the dots''), to get
\be
\frac{\pl L}{\pl {\dot q}^i} = \Sigma_j \; \frac{\pl L}{\pl {\dot q}'^j}\frac{\pl 
{\dot q}'^j}{\pl {\dot q}^i} 
= \Sigma_j \; \frac{\pl L}{\pl {\dot q}'^j}\frac{\pl q'^j}{\pl q^i} \;\; .
\ee
Then using the transformation law for components of a vector field
\be
X'^i = \Sigma_j \; \frac{\pl q'^i}{\pl q^j} X^j .
\label{trsfmvecfield}
\ee
and relabelling $i$ and $j$, we deduce:
\be
p'_X = \Sigma_i \; X'^i \frac{\pl L}{\pl {\dot q}'^i} = \Sigma_{ij} \;
 X^j \frac{\pl q'^i}{\pl q^j}\frac{\pl L}{\pl {\dot q}'^i}
 = \Sigma_{ij} \;
 X^i \frac{\pl q'^j}{\pl q^i}\frac{\pl L}{\pl {\dot q}'^j}
 = \Sigma_i \; X^i \frac{\pl L}{\pl {\dot q}^i} \equiv p_X \; .
\ee 
Finally,  I remark incidentally that in the geometric formulation of Lagrangian 
mechanics (Section \ref{geomperspLag}) , the coordinate-independence of $p_X$ 
becomes, unsurprisingly, a triviality. Namely: $p_X$ is obviously the contraction 
of $X$ as lifted to $TQ$ with the canonical  1-form on $TQ$ that we defined in eq. 
\ref{defpdq}:
\be
\theta_L := \frac{\pl L}{\pl {\dot q}^i}dq^i \;\;.
\label{defpdq2nd}
\ee
We will return to this at the end of Section \ref{NoetherLagGeomSubsub}.

\subsection{Noether's theorem; and examples}\label{Noetsubsubsec}
Given just the definition of conjugate momentum, eq. \ref{defconjugmommX}, the 
proof of Noether's theorem is  immediate. (The interpretation and properties of 
this momentum, discussed in the last Subsection, are not needed.)
The theorem says:
\begin{quote}
{\bf {Noether's theorem for Lagrangian mechanics}} If $X$ is a (variational) 
symmetry of a system with Lagrangian $L(q,{\dot q},t)$, then $X$'s conjugate 
momentum is a constant of the motion.
\end{quote}
{\em Proof}: We just calculate the derivative of the momentum eq. 
\ref{defconjugmommX} along the solution curves in $TQ$, and apply Lagrange's 
equations and the definitions of $Y^i$, and of symmetry eq. 
\ref{definevarnlsymy2}:
\begin{eqnarray}
\frac{d p}{dt} = \Sigma_i \; \frac{d X^i}{dt}\frac{\pl L}{\pl {\dot q}^i} \; + \;
\Sigma_i \;  X^i \frac{d }{dt}\left(\frac{\pl L}{\pl {\dot q}^i}\right) \\ 
\nonumber
= \; \Sigma_i \; Y^i\frac{\pl L}{\pl {\dot q}^i} \; + \; \Sigma_i \; X^i \frac{\pl 
L}{\pl q^i} \;\; = 0 \;\;. \; \; \; {\rm{QED.}}
\label{provenoet}
\end{eqnarray}

{\em Examples}:--- This proof, though neat, is a bit abstract! So here are two 
examples, both of which return us to examples we have already seen.

\indent (1): The first example is a shift in a cyclic coordinate $q^n$: i.e. the 
case with which our discussion of Noether's theorem began at the end of  Section 
\ref{Lageqsec}. So suppose $q^n$ is cyclic, and define a vector field $X$ by
\be
X^1 = 0, \dots, X^{n-1} = 0, \; X^n = 1 \; .
\label{Xrectified}
\ee
So the displacements generated by $X$ are translations by an amount $\epsilon$ in 
the $q^n$-direction. Then $Y^i := \frac{d X^i}{dt}$ vanishes, and the definition 
of (variational) symmetry eq. \ref{definevarnlsymy2} reduces to
\be
\frac{\pl L}{\pl q^n} = 0 \; .
\label{qncyclicinnoet}
\ee
So since $q^n$ is assumed to be cyclic, $X$ is a symmetry. And the momentum 
conjugate to $X$, which Noether's theorem tells us is a constant of the motion, is 
the familiar one:  
\be
p_X := \Sigma_i \; X^i \frac{\pl L}{\pl {\dot q}^i} =  \frac{\pl L}{\pl {\dot 
q}^n} \;\; .
\label{pXispn!}
\ee

As mentioned in Section \ref{NoetherPreamble},  this example is {\em universal}, 
in that every symmetry $X$ arises, around any point where $X$ is non-zero,  from a 
cyclic coordinate in some local system of coordinates. This follows from the basic 
theorem about the local existence and uniqueness of solutions of ordinary 
differential equations. We can state the theorem as follows; (cf. e.g. Arnold 
(1973: 48-49, 77-78, 249-250), Olver (2000: Prop 1.29)).

Consider a system of $n$ first-order ordinary differential equations on an open 
subset $U$ of an $n$-dimensional manifold 
\be
{\dot {q^i}} = X^i(q) \equiv X^i(q^1,...,q^n) \;\; , \;\; q \in U \;\; ;
\label{arnold10de}
\ee 
equivalently, a vector field $X$ on $U$. Let $q_0$ be a non-singular point of the 
vector  field, i.e. $X(q_0) \neq 0$. Then in a sufficiently small neighbourhood 
$V$ of $q_0$, there is a coordinate system (formally, a diffeomorphism $f:V 
\rightarrow W \subset \mathR^n$) such that, writing $y_i: \mathR^n \rightarrow 
\mathR$ for the standard coordinates on $W$ and ${\bf e}_i$ for the $i$th standard 
basic vector of $\mathR^n$, eq. \ref{arnold10de} goes into the very simple form
\be
{\dot {\bf y}} ={\bf e}_n ; {\rm{\;\; i.e. \;\;}} 
{\dot y_n} = 1, \;\; {\dot y}_1 = {\dot y}_2 = \dots = {\dot y}_{n-1} = 0 
{\rm{\;\; in \;\;}} W \;\; .
\label{arnold10deparallelized}
\ee 
(In  terms of the tangent map (also known as: push-forward)  $f_*$ on tangent 
vectors that is induced by $f$: $f_*(X) = {\bf e}_n$ in $W$.) On account of eq. 
\ref{arnold10deparallelized}'s simple form, Arnold suggests the theorem might well 
be called the `rectification theorem'.

We should note two points about the theorem:\\
\indent (i): The rectifying coordinate system $f$ may of course be very hard to 
find. So the theorem by no means makes all problems ``trivially soluble''; cf. 
again footnote 4.\\
\indent (ii): The theorem has an immediate corollary about {\em local} constants 
of the motion. Namely: $n$ first-order ordinary differential equations have, 
locally, $n-1$ functionally independent constants of the motion (also known as: 
first integrals). They are given, in the above notation, by  $y_1,\dots,y_{n-1}$.

We now apply the rectification theorem, so as to reverse the reasoning in the 
above example of $q^n$ cyclic. That is: assuming $X$ is a symmetry, let us rectify 
it---i.e. let us pass to a coordinate system $(q)$ such that eq. \ref{Xrectified} 
holds. Then, as above, $Y^i := \frac{d X^i}{dt}$ vanishes; and $X$'s being a 
(variational) symmetry, eq. \ref{definevarnlsymy2}, reduces to $q^n$ being cyclic; 
and the momentum conjugate to $X$, $p_X$ reduces to the familiar conjugate 
momentum $p_n = \frac{\pl L}{\pl {\dot q}^n}$. Thus every symmetry $X$ arises 
locally from a cyclic coordinate $q^n$ and the corresponding conserved momentum is 
$p_n$. (But note that this may hold only ``very locally'': the domain $V$ of the 
coordinate system $f$ in which $X$ generates displacements in the direction of the  
cyclic coordinate $q^n$ can be smaller than the set $U$ on which $X$ is a 
symmetry.) 

In Section \ref{noetHamsimple}, the fact that every symmetry arises locally from a 
cyclic coordinate will be important for understanding the Hamiltonian version of 
Noether's theorem. 

(2): Let us now look at our previous example, the angular momentum of a free 
particle (eq. \ref{LagSpherl}), in the {\em cartesian} coordinate system, i.e. a 
coordinate system {\em without} cyclic coordinates. So let $q_1 := x, q_2 := y, 
q_3 := z$. (In this example, subscripts will again be a bit clearer.) Then a small 
rotation about the $x$-axis
\be
\dd x = 0,  \;\; \dd y = - \epsilon z, \;\; \dd z = \epsilon y
\ee
corresponds to a vector field $X$ with components 
\be
X_1 = 0, \;\; X_2 = - q_3, \;\; X_3 = q_2
\label{infmlrotnx}
\ee
so that the $Y_i$ are
\be
Y_1 = 0, \;\; Y_2 = - {\dot q}_3, \;\; X_3 = {\dot q}_2 \;\; .
\ee
For the Lagrangian
\be
L = \frac{1}{2}m({\dot q}^2_1 + {\dot q}^2_2 + {\dot q}^2_3)
\ee
$X$ is a (variational) symmetry since the definition  of symmetry eq. 
\ref{definevarnlsymy2} now reduces to
\be
\Sigma_i \; X_i \frac{\pl L}{\pl q_i} \; + \; \Sigma_i \; Y_i \frac{\pl L}{\pl 
{\dot q_i}}  = - {\dot q}_3 \frac{\pl L}{\pl {\dot q}_2} + {\dot q}_2 \frac{\pl 
L}{\pl {\dot q}_3} = 0 \; .
\ee
So Noether's theorem then tells us that $X$'s conjugate momentum is
\be
p_X := \Sigma_i \; X_i \frac{\pl L}{\pl {\dot q}_i} = X_2\frac{\pl L}{\pl {\dot 
q}_2} + X_3\frac{\pl L}{\pl {\dot q}_3} 
= -mz{\dot y} + my{\dot z}  
\ee 
which is indeed the $x$-component of angular momentum.

\subsubsection{A geometrical formulation}\label{NoetherLagGeomSubsub}
We can give a geometric formulation of Noether's theorem by using the vanishing of 
the Lie derivative  to express constancy along the integral  curves of a vector 
field. There are two vector fields on $TQ$ to consider: the dynamical vector field 
$D$ (cf. eq. \ref{LagDelta}), and the lift to $TQ$ of the vector field $X$ that is 
the variational symmetry.

 I will now write $\bar X$ for this lift. So given the vector  field $X$ on $Q$
\be
X = X^i(q)\frac{\pl }{\pl q^i} \;\; ,
\ee 
the lift $\bar X$ of $X$ to $TQ$ is, by eq. \ref{defineY},
\be
{\bar X} = X^i(q)\frac{\pl }{\pl q^i}  + \frac{\pl X^i(q)}{\pl q^j} {\dot q}^j 
\frac{\pl }{\pl {\dot q}^i} \;\; , 
\ee 
where the $q$ argument of $X^i$ emphasises that the $X^i$ do not depend on ${\dot 
q}$.

That $X$ is a variational symmetry means that in  $TQ$, the Lie derivative of $L$ 
along the lift $\bar X$ vanishes: ${\cal L}_{\bar X} L = 0$. On the other hand, we 
know from eq. \ref{defpdq2nd} that the momentum $p_X$ conjugate to $X$ is the 
contraction $< ; >$ of $\bar X$ with the canonical 1-form $\theta_L := \frac{\pl 
L}{\pl {\dot q}^i} dq^i$ on $TQ$:
\be
p_X \; := \; X^i \frac{\pl L}{\pl {\dot q}^i} \; \equiv \; < {\bar X}; \theta_L > 
\; .
\label{pXascontraction}
\ee    
So Noether's theorem says: 
\begin{center}
If ${\cal L}_{\bar X} L = 0$, then ${\cal L}_D < {\bar X} ; \theta_L > = 0$.
\end{center}

Note finally that eq. \ref{pXascontraction} shows that the theorem has no 
converse. That is: given that a dynamical variable $p:TQ \rightarrow \mathR$ is a 
constant of the motion, ${\cal L}_D p = 0$, there is no single vector field $\bar 
X$ on $TQ$ such that $p = < {\bar X} ; \theta_L >$. For given such a $\bar X$, one 
could get another by adding any field $\bar Y$ for which $< {\bar Y} ; \theta_L > 
= 0$. However, we will see in Section \ref{hamnvfs} that in Hamiltonian mechanics 
a constant of the motion {\em does} determine a corresponding vector field on the 
state space. 

\section{Hamiltonian  mechanics introduced}\label{Hammechs}

\subsection{Preamble}\label{Hammechspreamb}
From now on this paper adopts the Hamiltonian framework. As we shall see, its 
description of symmetry and conserved quantities is in various ways more 
straightforward and powerful than that of the Lagrangian framework. 

The main idea is to replace the ${\dot q}$s by the canonical momenta, the $p$s. 
More generally, the state-space is no longer the tangent bundle $TQ$ but a phase 
space $\Gamma$, which we take to be the cotangent bundle $T^*Q$. (Here, the phrase 
`we take to be' just signals the fact that eventually, in Section 
\ref{Glimpsegenl}, we will glimpse a more general kind of Hamiltonian state-space, 
viz. Poisson manifolds.)\\
\indent Admittedly, the theory on $TQ$ given by Lagrange's equations eq. 
\ref{eqn;lag} is equivalent to the Hamiltonian  theory on $T^*Q$ given by eq. 
\ref{HamFromLagSimple} below, once we assume the Hessian condition eq. 
\ref{nonzerohessian}.\\
\indent But of course, theories can be formally equivalent, but different as 
regards their power for solving problems, their heuristic value and even their 
interpretation. In our case, two advantages of Hamiltonian mechanics over 
Lagrangian mechanics  are commonly emphasised.  (i): The first concerns its 
greater power or flexibility  for describing a given system, that Lagrangian 
methods can also describe (and so its greater power for solving problems about 
such a system). (ii): The second concerns the broader idea of describing other 
systems. In more detail:---

\indent (i): Hamiltonian mechanics  replaces the group of {\em point 
transformations}, $q \rightarrow q'$ on $Q$, together with their lifts to $TQ$, by 
a ``corresponding larger'' group of transformations on $\Gamma$, the group of {\em 
canonical transformations} (also known as, for the standard case where $\Gamma = 
T^*Q$: the {\em symplectic group}).\\
\indent This group ``corresponds'' to the point transformations (and their lifts) 
in that while for any Lagrangian  $L$, Lagrange's equations eq. \ref{eqn;lag} are 
covariant  under all the point transformations, Hamilton's equations eq. 
\ref{HamFromLagSimple} below are (for any Hamiltonian $H$) covariant under all 
canonical transformations. And it     is a ``larger'' group because:\\
\indent (a) any point transformation together with its lift to $TQ$ is a canonical 
transformation: (more precisely: it naturally defines a canonical transformation 
on $T^*Q$);\\
\indent (b) not every canonical transformation is thus induced by a point 
transformation; for  a canonical transformation can ``mix'' the $q$s and $p$s in a 
way that point transformations and their lifts cannot.\\
\indent There is a rich and multi-faceted theory of canonical transformations, to 
which there are three main approaches---generating functions, integral invariants 
and symplectic geometry. I will adopt the symplectic approach, but not need many 
details about it. In particular, we will need only a few details about how the 
``larger'' group of canonical transformations makes for a more powerful version of 
Noether's theorem.

\indent (ii): The Hamiltonian framework  connects analytical mechanics with other 
fields of physics, especially statistical mechanics and optics. The first 
connection goes via canonical transformations, especially using the integral 
invariants approach. The second connection goes via Hamilton-Jacobi theory; (for  
a philosopher's exposition, with an eye on quantum theory, cf. Butterfield (2004b: 
especially Sections 7-9)).\footnote{Of course, some aspects of Hamiltonian 
mechanics  illustrate both (i) and (ii). For example, Liouville's theorem on the 
preservation of phase space volume illustrates both (i)'s integral invariants 
approach to canonical transformations and (ii)'s connection to statistical 
mechanics.} 

\indent With its theme of symmetry and conservation, this paper will illustrate 
(i), greater power in describing a given system, rather than (ii), describing 
other systems. As to (i), we will see two main ways in which the Hamiltonian 
framework is more powerful than the Lagrangian one. First, cyclic coordinates will 
``do more work for us'' (Section \ref{Hameq}). Second, the Hamiltonian version of 
Noether's theorem is both: more powerful, thanks to the use of the ``larger'' 
group of canonical transformations; and more easily proven, thanks to the use of 
Poisson brackets (Section \ref{PoissNoet}). 

So from now on, the broad plan is as follows. After Section \ref{Hameq}'s 
deduction of Hamilton's equations, Section \ref{sympformintro} introduces  
symplectic structure, starting from the ``naive'' form of the symplectic matrix. 
Section \ref{PoissNoet} presents Poisson brackets, and the Hamiltonian version of 
Noether's theorem. Finally, Section \ref{geomperspHam} gives a geometric 
perspective, corresponding to Section \ref{geomperspLag}'s geometric perspective 
on the Lagrangian framework.

\subsection{Hamilton's equations}\label{Hameq}
\subsubsection{The equations introduced}\label{Hameqintrodd}
Recall the vision in (5) of Section \ref{tgtble}: that we seek $2n$ new variables, 
$\xi^{\al}$ say, $\al = 1,..., 2n$ in which Lagrange's equations take the simple 
form
\be
\frac{d \xi^{\al}}{dt} = f_{\al}(\xi^1,...\xi^{2n}) \; .
\label{simpledesire}
\ee
We can find the desired variables $\xi^{\al}$ by using the canonical momenta 
\be
p_i := \frac{\pl L}{\pl {\dot q}^i} =: L_{{\dot q}^i} \; ,
\label{definepsubifromqsupi}
\ee  
to write the $2n$ Lagrange equations as
\be
\frac{d p_i}{dt} = \frac{\pl L}{d q^i} \;\;\;  ; \; \;\; \frac{d q^i}{dt} = {\dot 
q}^i \;.
\label{LagAsSimple}
\ee 
These are of the desired simple form, except that the right hand sides need to be 
written as functions of $(q,p,t)$ rather than $(q,{\dot q},t)$. (Here and in the 
next two paragraphs, we temporarily allow time-dependence, since the deduction is 
unaffected: the time variable is ``carried along unaffected''. In the terms of 
Section \ref{Lageqsec}, this means allowing non-scleronomous constraints and a 
time-dependent work-function $U$.) 

For the second group of $n$ equations, this is in principle straightforward, given 
our assumption of a non-zero Hessian, eq. \ref{nonzerohessian}. This implies that 
we can invert eq. \ref{definepsubifromqsupi} so as to get the $n$ ${\dot q}^i$ as 
functions of $(q,p,t)$. We can then apply this to the first group of equations; 
i.e. we substitute ${\dot q}^i(q,p,t)$ wherever ${\dot q}^i$ appears in any right 
hand side $\frac{\pl L}{d q^i}$.

But we need to be careful: the partial derivative of $L(q,{\dot q},t)$ with 
respect to $q^i$ is not the same as the partial derivative of ${\hat L}(q,p,t) := 
L(q,{\dot q}(q,p,t),t)$ with respect to $q^i$, since the first holds fixed the 
${\dot q}$s, while the second holds fixed the $p$s. A comparison of these partial 
derivatives leads, with algebra, to the result that if we define the {\em 
Hamiltonian function} by
\be
H(q,p,t) := p_i{\dot q}^i(q,p,t) - {\hat L}(q,p,t)
\label{defineHamJS}
\ee
then the $2n$ equations eq. \ref{LagAsSimple} go over to {\em Hamilton's 
equations}
\be
\frac{d p_i}{dt} = - \frac{\pl H}{\pl q^i} \;\;\; ; \;\;\; \frac{d q^i}{dt} = 
\frac{\pl H}{\pl p_i} \;.
\label{HamFromLagSimple}
\ee
So we have cast our $2n$ equations in the simple form, $\frac{d \xi^{\al}}{dt} = 
f_{\al}(\xi^1,...\xi^{2n})$, requested in (5) of Section \ref{geomperspLag}. More 
explicitly: defining
\be
\xi^{\al} = q^{\al}, \;\; {\al} = 1,...,n \;\;\; ; \;\;\; \xi^{\al} = p_{{\al}- 
n}, \;\; {\al} = n+1,...,2n
\label{definexisupal} 
\ee
Hamilton's equations become
\be
{\dot \xi}^{\al} = \frac{\pl H}{\pl \xi^{\al + n}}, \;\; {\al} = 1,...,n\;\;\; ; 
\;\;\; {\dot \xi}^{\al} =  - \frac{\pl H}{\pl \xi^{\al - n}}, \;\; {\al} = 
n+1,...,2n \;\; .
\label{Hamxisupal} 
\ee
To sum up: a single function $H$ determines, through its partial derivatives, the 
evolution of all the $q$s and $p$s---and so, the evolution of the state of the 
system.

\subsubsection{Cyclic coordinates in the Hamiltonian framework}\label{cyclicinHam}
Just from the form of Hamilton's equations, we can immediately see a result that 
is significant for our theme of how symmetries and conserved quantities reduce the 
number of variables involved in a problem. In short, we can see that with 
Hamilton's equations in hand, cyclic coordinates will ``do more work for us'' than 
they do in the Lagrangian framework.\\
\indent More specifically, recall the basic Lagrangian result from the end of 
Section \ref{Lageqsec}, that the generalized momentum $p_n := \frac{\pl L}{\pl 
{\dot q}^n}$ is conserved if, indeed iff, its conjugate coordinate $q^n$ is 
cyclic, $\frac{\pl L}{\pl q^n} = 0$. And recall from Section \ref{Noetsubsubsec} 
that this result underpinned Noether's theorem in the precise sense of being 
``universal'' for it. Corresponding results hold in the Hamiltonian  
framework---but are in certain ways more powerful.

\indent Thus we first observe that the transformation ``from the $\dot q$s to the 
$p$s'', i.e. the transition between Lagrangian and Hamiltonian frameworks, does 
not involve the dependence on the $q$s.  More precisely: partially differentiating 
eq. \ref{defineHamJS} with respect to $q^n$, we obtain
\be
\frac{\pl H}{\pl q^n} \equiv \frac{\pl H}{\pl q^n}\mid_{p; q^i, i \neq n} \;\; = 
\;\; - \frac{\pl L}{\pl q^n} \equiv \frac{\pl L}{\pl q^n}\mid_{{\dot q}; q^i, i 
\neq n} \; .
\label{Hqn=Lqn}
\ee 
(The other two terms are plus and minus $p_i \frac{\pl {\dot q}^i}{\pl q^n}$, and 
so cancel.) So a coordinate $q^n$ that is cyclic in the Lagrangian sense is also 
cyclic in the obvious Hamiltonian sense, viz. that $\frac{\pl H}{\pl q^n} = 0$. 
But by Hamilton's equations, this is equivalent to ${\dot p}_n = 0$. So we have 
the result corresponding to the Lagrangian  one: $p_n$ is conserved iff $q_n$ is 
cyclic (in the Hamiltonian sense).

We will see in Section \ref{noetHamsimple} that this result underpins the 
Hamiltonian version of Noether's theorem; just as the corresponding  Lagrangian 
result underpinned the Lagrangian version of Noether's theorem (cf. discussion 
after eq. \ref{pXispn!}). 

But we can already see that this result gives the Hamiltonian formalism  an 
advantage  over the Lagrangian. In the latter, the generalized velocity 
corresponding to a cyclic coordinate, $q_n$ will  in general still occur in the 
Lagrangian. The Lagrangian will be $L(q_1,\dots,q_{n-1},{\dot q}_1,\dots,{\dot 
q}_n, t)$, so that we still face a problem in $n$ variables.\\
\indent But in the Hamiltonian formalism, $p_n$ will be a constant of the motion, 
$\alpha$ say, so that the Hamiltonian will be 
$H(q_1,\dots,q_{n-1},p_1,\dots,p_{n-1},\alpha,t)$. So we now face a problem in $n 
- 1$ variables, $\alpha$ being simply determined by the initial conditions. That 
is: after solving the problem in $n - 1$ variables, $q_n$ is determined just by 
quadrature: i.e. just by integrating (perhaps numerically) the equation
\be
{\dot q}_n = \frac{\pl H}{\pl \alpha} \; ,
\label{qnbyquadrat}
\ee
where, thanks to having solved the problem in $n-1$ variables, the right-hand side 
is now an explicit function of $t$.

This result is very simple. But it is an important illustration of the power of 
the Hamiltonian framework. Indeed, Arnold remarks (1989: 68) that `almost all the 
solved problems in mechanics have been solved by means of' it!\\
\indent No doubt his point is, at least in part, that  this result underpins the 
Hamiltonian version of Noether's theorem. But I should add that the result also  
motivates the study of 
various notions related to the idea of cyclic coordinates, such as constants of 
the motion being in involution (i.e. having zero Poisson bracket with each other), 
and a system being completely integrable (in the sense of Liouville). These 
notions have played a large part in the way that Hamiltonian mechanics has 
developed, especially in its theory of canonical transformations. And they play a 
large part in the way Hamiltonian mechanics  has solved countless problems. But as 
announced in Section \ref{Hammechspreamb}, this paper will not go into these 
aspects of Hamiltonian mechanics, since they are not needed for our theme of 
symmetry and conservation; (for a philosophical discussion of these aspects, cf. 
Butterfield 2005). 

\subsubsection{The Legendre transformation and variational 
principles}\label{Legtrsfmnvarnal}
To end this Subsection, I note two aspects of this transition from Lagrange's 
equations to Hamilton's. For, although I shall not need details about them, they 
each lead to a rich theory:\\
\indent (i): The transformation ``from the $\dot q$s to the $p$s'' is the {\em 
Legendre transformation}. It has a striking geometric  interpretation. In the 
simplest case, it concerns the fact that one can describe a smooth convex real 
function $y =f(x), f''(x) > 0$, not by the pairs of its arguments and values 
$(x,y)$, but by the pairs of its gradients at points $(x,y)$ and the intercepts of 
its tangent lines with the $y$-axis. Given the non-zero Hessian (eq. 
\ref{nonzerohessian}), one readily proves various results: e.g. that the geometric  
interpretation extends to higher dimensions, and that the transformation is 
self-inverse, i.e. its square is the identity. For details, cf. e.g.: Arnold 
(1989: Chapters 3.14, 9.45.C), Courant and Hilbert (1953: Chapter IV.9.3; 1962, 
Chapter I.6), Jos\'{e} and Saletan (1998: 212-217), Lanczos (1986: Chapter 
VI.1-4). The Legendre transformation is also described using modern geometry's 
idea of a {\em fibre derivative}; as we will see briefly in Section 
\ref{geomlegetrsfmn}.\\
\indent (ii): The transition to Hamilton's equations has achieved  more than we 
initially sought with our eq. \ref{simpledesire}. Namely: all the $f_{\al}$, all 
the right hand sides in Hamilton's equations, are up to a sign, partial 
derivatives of a single function $H$. In the Hamiltonian framework, it is 
precisely this feature that underpins the possibility of expressing the equations 
of motion by variational principles; (of course, the Lagrangian framework has a 
corresponding feature). But as I mentioned, this paper does not discuss 
variational principles; for details cf. e.g. Lanczos (1986: Chapter VI.4) and 
Butterfield (2004: especially Section 5.2).

To sum up this introduction to Hamilton's equations:--- Even once we set aside (i) 
and (ii), these equations  mark the beginning of a rich and multi-faceted theory. 
At the centre lies the $2n$-dimensional phase space $\Gamma$ coordinatized by the 
$q$s and $p$s: or more precisely, as we shall see later, the cotangent bundle 
$T^*Q$. The structure of Hamiltonian mechanics is encoded in the structure of 
$\Gamma$, and thereby in the coordinate transformations on $\Gamma$ that preserve 
this structure, especially the form of Hamilton's equations: the canonical 
transformations. As I mentioned in Section \ref{Hammechspreamb},  these 
transformations can be studied from three main perspectives: generating functions, 
integral invariants and symplectic structure---but I shall only need the last.

\subsection{Symplectic  forms on vector spaces}\label{sympformintro}
I shall introduce symplectic structure by giving Hamilton's equations a yet more 
symmetric appearance. This will lead to some elementary  ideas about area in 
$\mathR^m$ and symplectic  forms on vector spaces: ideas which will later be 
``made local'' by taking the relevant copy of $\mathR^m$ to be the tangent space 
at a point of a manifold. (As usually formulated, Hamiltonian mechanics is 
especially concerned with the case $m = 2n$.)
 
\subsubsection{Time-evolution from the gradient of $H$}\label{DetsysevolngradH}
  Writing $\bf 1$ and $\bf 0$ for the $n \times n$ identity and zero matrices 
respectively, we define the $2n \times 2n$ {\em symplectic matrix} ${\o}$ by
\be
\o : = \left(\begin{array}{cc}
{\bf 0} & {\bf 1} \\
{\bf {-1}} & {\bf 0} \\
\end{array}\right) \;\; .
\label{eq;defineOmega}
\ee
$\o$ is antisymmetric, and has the properties, writing $\; \tilde{} \;$ for the 
transpose of a matrix, that
\be
{\tilde {\o}} = - {\o} = {\o}^{-1} \;\;\; \mbox{ so that } \;\;\;{\o}^2 = - {\bf 
1} \;\;\; \mbox{ ; } \;\; \mbox{ also } \;\;\; {\rm det}\;{\o} = 1.
\label{eq;propiesOmega}
\ee
Using $\o$, Hamilton's equations eq. \ref{Hamxisupal} get the more symmetric form, 
in matrix notation
\be
{\dot {\bf \xi}} = {\o}\frac{\pl H}{\pl {\bf \xi}} \;\; .
\label{eq;hamwithOmatrixnotation}
\ee
In terms of components,  writing $\o^{{\al\bb}}$ for the matrix elements of $\o$, 
and defining $\pl_{\al} := \pl \;/ \pl \xi^{\al}$,  eq. \ref{Hamxisupal} become
\be
{\dot \xi}^{\al} = \o^{\al\bb}\pl_{\bb}H .
\label{HamxisupalWithOmega} 
\ee
Eq. \ref{eq;hamwithOmatrixnotation} and \ref{HamxisupalWithOmega} show how $\o$ 
forms, from the naive gradient (column vector) $\nabla H$ of $H$ on the phase 
space $\Gamma$ of $q$s and $p$s, the vector field on $\Gamma$ that gives the 
system's evolution: the {\em Hamiltonian vector field}, often written $X_H$. At a 
point $z = (q,p) \in \Gamma$, eq. \ref{eq;hamwithOmatrixnotation} can be written
\be
X_H (z) = \o \nabla H(z).
\label{introX_H}
\ee 
The vector field $X_H$ is also written as $D$ (for `dynamics'), on analogy with 
the Lagrangian  framework's vector field $D$ of eq. \ref{LagDelta} in Section 
\ref{geomperspLag}.

In Section \ref{geomperspHam}, we will see how this definition of a {\em vector} 
field from a gradient, i.e. a {\em covector} or 1-form field, arises from 
$\Gamma$'s being a cotangent bundle. More precisely, we will see that any 
cotangent bundle has an intrinsic symplectic structure that provides, at each 
point of the base-manifold, a natural i.e. basis-independent isomorphism between 
the tangent space   and the cotangent space. For the moment, we:\\
\indent (i) note a geometric interpretation of $\o$ in terms of area (Section 
\ref{interpareas}); and then\\
\indent (ii) generalize the above discussion of $\o$ into the definition  of a 
symplectic form for a fixed vector  space (Section \ref{formsassoclin}).

\subsubsection{Interpretation in terms of areas}\label{interpareas}
Let us begin with the simplest possible case: $\mathR^2 \ni (q,p)$, representing 
the phase  space of a particle constrained to one spatial dimension. Here, the $2 
\times 2$ matrix
\be
\o : = \left(\begin{array}{cc}
0 & 1 \\
-1 & 0 \\
\end{array}\right) \;\;
\label{eq;defineOmega2times2}
\ee
defines the antisymmetric bilinear form on $\mathR^2$:
\be
A: ((q^1,p_1),(q^2,p_2)) \in \mathR^2 \times \mathR^2 \mapsto q^1p_2 - q^2p_1  \in 
\mathR
\label{defineAR2}
\ee
since
\be
q^1p_2 - q^2p_1 = \left(\begin{array}{cc}
q^1 & p_1 \\
\end{array}\right) 
\left(\begin{array}{cc}
0 & 1 \\
-1 & 0 \\
\end{array}\right)
\left(\begin{array}{c}
q^2 \\
p_2 \\
\end{array}\right)
 = {\rm det}\; \left(\begin{array}{cc}
q^1 & q^2 \\
p_1 & p_2 \\
\end{array}\right) \;\; .
\label{explainAR2}
\ee
It is easy to prove that $A((q^1,p_1),(q^2,p_2)) \equiv q^1p_2 - q^2p_1$
is the signed area of the parallelogram spanned by $(q^1,p_1), (q^2,p_2)$, where 
the sign is positive (negative) if the shortest rotation from $(q^1,p_1)$ to 
$(q^2,p_2)$ is anti-clockwise (clockwise).

Similarly in $\mathR^{2n}$: the matrix $\o$ of eq. \ref{eq;defineOmega} defines an 
antisymmetric bilinear form on $\mathR^{2n}$ whose value on a pair $(q,p) \equiv 
(q^1,...q^n; p_1,...,p_n), (q',p') \equiv (q'^1,...q'^n; p'_1,...,p'_n)$ is the 
sum of the signed areas of the $n$ parallelograms formed by the projections of the 
vectors $(q,p), (q',p')$ onto the $n$ pairs of coordinate planes labelled 
$1,...,n$. That is to say, the value is:
\be
\Sigma^n_{i=1} \; q^i p'_{i} - q'^{i} p_i \;\;.
\label{explainARn}
\ee

This induction of bilinear forms from antisymmetric matrices can be generalized: 
there is a one-to-one correspondence between forms and matrices. In more detail:  
there is a one-to-one correspondence between antisymmetric bilinear forms on 
$\mathR^2$ and antisymmetric $2 \times 2$ matrices. It is easy to check that any 
such form, $\o$ say, is given, for any basis $v, w$ of $\mathR^2$, by the matrix 
$\left(\begin{array}{cc}
0 & \o(v,w) \\
-\o(v,w) & 0 \\
\end{array}\right)$. Similarly for any integer $n$: one easily shows that  there 
is a one-to-one correspondence between antisymmetric bilinear forms on $\mathR^n$ 
and antisymmetric $n \times n$ matrices. (In Hamiltonian mechanics as usually 
formulated, we consider the case where $n$ is even and the matrix is non-singular, 
as in eq. \ref{eq;defineOmega}. But when one generalizes to Poisson manifolds (cf. 
Section \ref{Glimpsegenl}) one allows $n$ to be odd, and the matrix to be 
singular.) 

This geometric interpretation of $\o$ is important for two reasons.\\
\indent (i): The first reason is that the idea of an antisymmetric bilinear form 
on a copy of $\mathR^{2n}$ is the main part of the definition of a symplectic 
form, which is the central notion in the usual geometric formulation of 
Hamiltonian mechanics. More details in Section \ref{formsassoclin}, for a fixed 
copy of  $\mathR^{2n}$; and in Section \ref{geomperspHam},  where   the form is 
defined on many copies of $\mathR^{2n}$, each copy being the tangent space at a 
point in the cotangent bundle $T^*Q$.

\indent (ii): The second reason is that the idea of (signed) area underpins the 
theory of forms (1-forms, 2-forms etc.): i.e. antisymmetric multilinear functions 
on products of copies of $\mathR^n$.
 And when these copies of $\mathR^n$ are copies of the tangent space at (one and 
the same) point in a manifold,  these forms lead to the whole theory of 
integration on manifolds. One needs this theory in order to make rigorous sense of 
any integration on a manifold beyond the most elementary (i.e. line-integrals); so 
it is  crucial for almost any mathematical or physical theory using manifolds. In 
particular, it is crucial for Hamiltonian mechanics. So no wonder the {\em 
maestro} says that `Hamiltonian mechanics cannot be understood without 
differential forms' (Arnold 1989, p. 163).

\indent However, it turns out that this paper will not need many details about 
forms and the theory of integration. This is essentially because we focus only on 
solving mechanical problems, and simplifying them by appeals to symmetry. This 
means we will focus on  line-integrals: viz. integrating with respect to time the 
equations of motion; or equivalently, integrating the dynamical vector field  on 
the state space. We have already seen this vector field as $X_H$ in eq. 
\ref{introX_H}; and we will see it again, for example in terms of Poisson brackets 
(eq. \ref{HamDelta0}), and in geometric terms  (Section \ref{geomperspHam}). But 
throughout, the main idea will be as suggested by eq. \ref{introX_H}: the vector 
field is determined by the symplectic matrix, ``at'' each point in the manifold 
$\Gamma$, acting on the gradient of the Hamiltonian function $H$.\\
\indent So in short: focussing on line-integrals enables us to side-step most of 
the theory of forms.\footnote{But forms  are essential for understanding 
integration over surfaces of dimension two or more: which one needs for the 
integral invariants approach to Hamiltonian mechanics, and its deep connection 
with Stokes' theorem.}

\subsubsection{Bilinear forms and associated linear maps}\label{formsassoclin}
We now generalize from the symplectic matrix $\o$ to a symplectic form; in five 
extended comments.
 
(1): {\em Preliminaries}:---\\
Let $V$ be a (real finite-dimensional) vector space, with basis 
$e_1,...,e_i,...e_n$. We write $V^*$ for the dual space, and $e^1,...,e^i,...e^n$ 
for the dual basis: $e^i(e_j) := \delta^i_j$. 

We recall that the isomorphism $e_i \mapsto e^i$ is basis-dependent: for a 
different basis, the corresponding isomorphism would be a different map. Only with 
the provision of appropriate extra structure would this isomorphism be 
basis-independent.\\
\indent For physicists, the most familiar example of such a structure is the 
spacetime metric ${\bf g}$ in relativity theory. In terms of components, this 
basis-independence shows up  in the way that ${\bf g}$ and its inverse lower and 
raise indices.  As we will see in a moment, the underlying mathematical  point is 
that because ${\bf g}$ is a  bilinear form on a vector space $V$, i.e. ${\bf g}: V 
\times V \rightarrow \mathR$, and is non-degenerate, any $v \in V$ defines, 
independently of any choice of basis, an element of $V^*$: viz. the map $u \in V 
\mapsto {\bf g}(u,v)$.  (In fact, $V$ is the tangent space at a spacetime point; 
but this physical interpretation is irrelevant to the mathematical argument.) We 
will also see that Hamiltonian mechanics has a non-degenerate bilinear form, viz. 
a symplectic form, that similarly gives a basis-independent isomorphism between a 
vector space and its dual. (Roughly speaking, this vector space will be the 
$2n$-dimensional space  of the $q$s and $p$s.)

 On the other hand: for any vector space $V$, the isomorphism between $V$ and 
$V^{**}$ given by
\be
e_i \mapsto [e_i] \in V^{**}: e^j \in V^* \mapsto e^j(e_i) = \delta^j_i 
\label{isodoubledual}
\ee 
is basis-independent, and so we identify $e_i$ with $[e_i]$, and $V$ with 
$V^{**}$. We will  write $< \; ; \; >$ (also written $< \; , \; >$) for the 
natural pairing (in either order) of $V$ and $V^*$: e.g. $<e_i \; ; \; e^j> \; = 
\; <e^j \; ; \; e_i> \; = \; \delta^j_i$.

A linear map $A:V \rightarrow W$ induces (basis-independently) a {\em transpose} 
(aka: dual), written $\tilde{A}$ (or $A^T$ or $A^*$), $\tilde{A}: W^* \rightarrow 
V^*$ by
\be
\forall \al \in W^*, \forall v \in V: \; \; {\tilde{A}}(\al)(v) \; \equiv \; 
<{\tilde{A}}(\al) \; ; \; v> \; := \; \al(A(v)) \; \equiv \;(\al \circ A)(v) \; .
\label{deftranspose}
\ee 

If $A:V \rightarrow W$ is a linear map between real finite-dimensional vector 
spaces, its matrix with respect to bases $e_1,...,e_i,...e_n$ and 
$f_1,...,f_j,...f_m$ of $V$ and $W$ is given by: 
\be
A(e_i) = A^j_i f_j \;\; ; {\rm{\;\;\;\; i.e. \;\; with \;\;}} v = v^i e_i, \;\; 
(A(v))^j = A^j_i v^i \;\;.
\label{mxAlinmap}
\ee
So the upper index labels rows, and the lower index labels columns. 
Similarly, if $A: V \times W \rightarrow \mathR$ is a bilinear form, its matrix 
for these bases is defined as
\be
A_{ij} := A(e_i,f_j)
\label{mxAbilform}
\ee
so that on vectors $v = v^i e_i, w = w^j f_j$, we have: $A(v,w) = v^i A_{ij} w^j.$

(2): {\em Associated maps and forms}:---\\
Given a bilinear form $A: V \times W \rightarrow \mathR$, we define the {\em 
associated  linear map} $A^{\fl}:V \rightarrow W^*$ by
\be
A^{\fl}(v)(w) \; := \;  A(v,w) \;\; .
\label{definebil'sassoclin}
\ee
Then $A^{\fl}(e_i) = A_{ij}f^j$: for both sides send any $w = w^j f_j$ to 
$A_{ij}w^j$. That is: the matrix of $A^{\fl}$ in the bases $e_i, f^j$ of $V$ and 
$W^*$ is $A_{ij}$:
\be
[A^{\fl}]_{ij} = A_{ij}.
\label{commonmx}
\ee

On the other hand, we can proceed from linear maps to associated bilinear forms. 
Given a linear map $B: V \rightarrow W^*$, we define the {\em associated bilinear 
form} $B^{\sh}$ on $V \times W^{**} \cong V \times W$ by
\be
B^{\sh}(v,w) \;\; = \;\; < B(v) \; ; \; w > \; .
\label{definemap'sassocbil}
\ee
If we put $A^{\fl}$ for $B$ in eq. \ref{definemap'sassocbil}, its associated 
bilinear form, acting on vectors $v = v^i e_i, w = w^j f_j$, yields, by eq. 
\ref{definebil'sassoclin}:
\be
(A^{\fl})^{\sh}(v,w) \;\; = \;\; < A^{\fl}(v) \; ; \; w > \;\; = \;\; A(v,w) \; .
\label{flatsharp}
\ee
One similarly shows that if $B:V \rightarrow W^*$, then $\forall  w \in W$:
\be
 (B^{\sh})^{\fl}(v)(w) \equiv <(B^{\sh})^{\fl}(v) \; ; \; w> \; = \; B(v)(w) 
\equiv <B(v) \; ; \; w> \; {\rm{\; so \; that \;}} \; (B^{\sh})^{\fl} = B \; .
\label{sharpflat}
\ee
So the flat and sharp operations, $^{\fl}$ and $^{\sh}$, are inverses.

(3): {\em Tensor products}:---\\
It will  sometimes be helpful to put the above ideas in terms of {\em tensor 
products}. If $v \in V, w \in W$, we can think of $v$ and $w$ as elements  of 
$V^{**}, W^{**}$ respectively. So we define their tensor product as a bilinear 
form on $V^* \times W^*$ by requiring for all $\al \in V^*, \bb \in W^*$:
\be
(v \otimes w)(\al,\bb) \;\; := \;\; v(\al)w(\bb) \; \equiv \; < v \; ; \; \al > < 
w \; ; \; \bb > \; .
\label{deftsrprod}
\ee 
Similarly for other choices of vector spaces or their duals. Given $\al \in V^*, 
\bb \in W^*$, their tensor product is a bilinear form on $V \times W$:
\be
(\al \otimes \bb)(v,w) \;\; := \;\; \al(v)\bb(w) \; \equiv \; < v \; ; \; \al > < 
w \; ; \; \bb > \; .
\label{deftsrprod1}
\ee 
Similarly, we can think of $\al \in V^*, w \in W$ as elements of $V^*$ and 
$W^{**}$ respectively, and so define their tensor product as a bilinear form on $V 
\times W^*$:
\be
(\al \otimes w)(v,\bb) \;\; := \;\; \al(v)w(\bb) \; \equiv \; < v \; ; \; \al > < 
w \; ; \; \bb > \; .
\label{deftsrprod2}
\ee 
In this way we can express the linear map $A: V \rightarrow W$ in terms of tensor 
products. Since 
\be
A(e_i) = A^j_i f_j \;\;\;\;  {\rm{iff}} \;\; <A(e_i) ; f^j > \;\;  = \; A^j_i 
\label{linmapfromtsrprod1}
\ee
eq. \ref{deftsrprod2} implies that
\be
A \; = \;  A^j_i \; e^i \otimes f_j \;\; .
\label{linmapfromtsrprod2}
\ee
Similarly, a bilinear form $A: V \times W \rightarrow \mathR$ with matrix $A_{ij} 
:= A(e_i,f_j)$ (cf. eq. \ref{mxAbilform}) is:
\be
A \; = \;  A_{ij} \; e^i \otimes f^j
\label{bilformfromtsrprod}
\ee
The definitions of tensor product eq. \ref{deftsrprod}, \ref{deftsrprod1} and 
\ref{deftsrprod2} generalize to higher-rank tensors (i.e. multilinear maps whose 
domains have more than two factors). But we will not need these generalizations. 

(4): {\em Antisymmetric and non-degenerate forms}:---\\
We now specialize to the forms and maps of central interest in Hamiltonian 
mechanics. We take $W = V$, dim($V$)=$n$, and define a bilinear form $\o: V \times 
V \rightarrow \mathR$ to be:\\
\indent (i): {\em antisymmetric} iff: $\o(v,v') = - \o(v,v')$;\\
\indent (ii): {\em non-degenerate} iff: if $\o(v,v') = 0 \;\; \forall v' \in V$, 
then $v = 0$.\\
The form $\o$ and its associated linear map $\o^{\fl}: V \rightarrow V^*$ now have 
a square matrix $\o_{ij}$ (cf. eq. \ref{commonmx}). We define the {\em rank} of 
$\o$ to be the rank of this matrix: equivalently, the dimension of the range 
$\o^{\fl}(V)$. 

We will also need the antisymmetrized version of eq. \ref{deftsrprod1} that is 
definable when $W = V$. Namely, we define  the {\em wedge-product} of $\al, \bb 
\in V^*$ to be the antisymmetric bilinear form on $V$, given by
\be
\al \wedge \bb: (v,w) \in V \times V \mapsto (\al(v))(\bb(w)) - (\al(w))(\bb(v)) 
\in \mathR \; .
\label{defwedgegeneral}
\ee  
(The connection with Section \ref{interpareas}, especially eq. \ref{explainARn}, 
will become clear in a moment; and will be developed in Section 
\ref{introduceforms}.A.) 
  
It is easy to show that for any  bilinear form $\o: V \times V \rightarrow 
\mathR$: $\o$ is non-degenerate iff the matrix $\o_{ij}$ is non-singular iff 
$\o^{\fl}: V \rightarrow V^*$ is an isomorphism.\\
\indent So a non-degenerate bilinear form establishes a basis-independent 
isomorphism between $V$ and $V^*$; cf. the discussion of the spacetime metric $\bf 
g$ in (1) at the start of this Subsection.\\
\indent Besides, this isomorphism $\o^{\fl}$ has an inverse, suggesting another 
use of the sharp notation, viz. $\o^{\sh}$ is defined to be $(\o^{\fl})^{-1}: V^* 
\rightarrow V$. The isomorphism  $\o^{\sh}: V^* \rightarrow V$ corresponds to 
$\o$'s role, emphasised in Section \ref{DetsysevolngradH}, of defining a vector 
field $X_H$ from $dH$. (But we will see in a moment that the space $V$ implicitly 
considered in Section \ref{DetsysevolngradH} had more structure than being just 
any finite-dimensional real vector space: viz. it was of the form $W \times W^*$.)

NB: This definition of  $\; ^{\sh}$ is of course {\em not} equivalent to our 
previous definition, in eq. \ref{definemap'sassocbil}, since:\\
\indent (i): on our previous definition, $^{\sh}$ carried a linear map to a 
bilinear form, which reversed the passage by $^{\fl}$ from bilinear  form to 
linear  map, in the sense  that for a bilinear form $\o$, we had $(\o^{\fl})^{\sh} 
= \o$; cf. eq. \ref{flatsharp};\\
\indent (ii): on the present definition, $^{\sh}$ carries a bilinear form $\o: V 
\times V \rightarrow \mathR$ to a linear map $\o^{\sh}: V^* \rightarrow V$, which 
inverts $^{\fl}$ in the sense ({\em different} from (i)) that
\be
\o^{\sh} \circ \o^{\fl} = id_V \;\; {\rm{and}} \;\; \o^{\fl} \circ \o^{\sh} = 
id_{V^*} \; .
\label{othersharpasinverse} 
\ee
So beware: though not equivalent, both definitions are used! But it is a natural 
ambiguity, in so far as the definitions ``mesh''. For example, one easily shows 
that our second definition, i.e.  eq. \ref{othersharpasinverse}, is  equivalent to 
a natural expression:
\be
\forall \al, \bb \in V^*: \; \; < \o^{\sh}(\al), \bb > \; := \; 
\o((\o^{\fl})^{-1}(\al), (\o^{\fl})^{-1}(\bb)) \; .
\ee

It is also straightforward to show that for any  bilinear form $\o: V \times V 
\rightarrow \mathR$: if $\o$ is antisymmetric of rank $r \leq n \equiv 
{\rm{dim}}(V)$, then $r$ is even. That is: $r = 2s$ for some integer $s$, and 
there is a basis $e_1,...,e_i,...,e_n$ of $V$ for which $\o$ has a simple 
expansion as wedge-products 
\be
\o \; = \; \Sigma^s_{i= 1} \; e^i \wedge e^{i+s} \;\; ;
\label{omegawiths}
\ee  
equivalently, $\o$ has the $n \times n$ matrix 
\be
\o \;  = \; \left(\begin{array}{ccc}
{\bf 0} & {\bf 1} & {\bf 0} \\
{\bf {-1}} & {\bf 0} & {\bf 0} \\
{\bf {0}} & {\bf 0} & {\bf 0} \\
\end{array}\right) \;\; .
\label{eq;gramschmidt}
\ee
where ${\bf 1}$ is the $s \times s$ identity matrix, and similarly for the zero 
matrices of various sizes. This {\em normal form} of antisymmetric bilinear forms 
is an analogue of the Gram-Schmidt theorem that an inner product space has an 
orthonormal basis, and is proved by an analogous argument.

(5): {\em Symplectic  forms}:---\\
As usually formulated, Hamiltonian mechanics  uses a non-degenerate antisymmetric  
bilinear form: i.e. $r = n$. So eq. \ref{eq;gramschmidt} loses its bottom row and 
right column consisting of zero matrices, and  reduces to the form of Section 
\ref{DetsysevolngradH}'s naive symplectic  matrix, eq. \ref{eq;defineOmega}. 
Equivalently: eq. \ref{omegawiths} reduces to eq. \ref{explainARn}. \\
\indent Accordingly, we define: a {\em symplectic form} on a (real 
finite-dimensional) vector space $Z$ is a non-degenerate antisymmetric bilinear 
form $\o$ on $Z$: $\o: Z \times Z \rightarrow \mathR$. $Z$ is then called a {\em 
symplectic vector space}. It follows that $Z$ is of even dimension. 

Besides, in Hamiltonian mechanics (as usually formulated) the vector space $Z$ is 
a product $V \times V^*$ of a vector space and its dual. Indeed, this was already 
suggested by:\\
\indent (i) the fact in (3) of Section \ref{tgtble}, that the canonical momenta 
$p_i := \frac{\pl L}{\pl {\dot q}^i}$ transform as a 1-form, and\\
\indent (ii) Section \ref{DetsysevolngradH}'s discussion of the one-form field 
$\nabla H$ determining a vector field $X_H$.\\
\indent Thus we define the {\em canonical symplectic form} $\o$ on $Z := V \times 
V^*$ by
\be
{\o}((v_1,\al_1), (v_2, \al_2)) \;\; := \;\; \al_2(v_1) - \al_1(v_2) \;\; .
\label{defcanlsympformvecsp}
\ee
So defined, $\o$ is by construction a symplectic form, and so has the normal form 
given by eq. \ref{eq;defineOmega}.

Given a symplectic vector space $(Z, \o)$, the natural question arises which 
linear maps $A:Z \rightarrow Z$ preserve the normal form given by eq. 
\ref{eq;defineOmega}. It is straightforward to show that this is equivalent to $A$ 
preserving the form of Hamilton's equations (for any Hamiltonian); so that these 
maps $A$ are called {\em canonical} (or {\em symplectic}, or {\em Poisson}). But 
since (as I announced) this paper does not need details about the theory of 
canonical transformations,  I will not go into details about this. Suffice it to 
say here the following.\\
\indent $A: Z \rightarrow Z$ is symplectic iff, writing $\; \tilde{} \;$ for the 
transpose (eq. \ref{deftranspose}) and using the second definition eq. 
\ref{othersharpasinverse} of $^{\sh}$, the following maps (both from $Z^*$ to $Z$) 
are equal:
\be
A \circ \omega^{\sh} \circ \tilde{A} = \omega^{\sh} \; \; ;
\label{charizesympA}
\ee
or in matrix notation, with the {\em matrix} $\omega$  given by eq. 
\ref{eq;defineOmega}, and again writing $\; \tilde{} \;$ for the transpose of a 
matrix
\be
A  \omega  \tilde{A} = \omega \; \; .
\label{charizesympAmatrix}
\ee
(Equivalent formulas are got by taking inverses. We get, respectively: $\tilde{A} 
\circ \omega^{\fl} \circ A = \omega^{\fl}$ and $\tilde{A}  \omega  A = \omega$.)\\
\indent The set of all such linear symplectic maps $A: Z \rightarrow Z$ form a 
group, {\em the symplectic group}, written Sp($Z,\o$). 

To sum up this Subsection:--- We have, for a vector space $V$, dim($V$) = $n$, and 
$Z := V \times V^*$:\\
\indent (i): the canonical symplectic form $\o: Z \times Z \rightarrow \mathR$; 
with normal form given by eq. \ref{eq;defineOmega};\\
\indent (ii): the associated linear map $\o^{\fl}: Z \rightarrow Z^*$; which is an 
isomorphism, since $\o$ is non-degenerate;\\
\indent (iii): the associated linear map $\o^{\sh}: Z^* \rightarrow Z$; which is 
an isomorphism, since $\o$ is non-degenerate; and is the inverse of $\o^{\fl}$; 
(cf. eq. \ref{othersharpasinverse}).\\
\indent We will see shortly that Hamiltonian mechanics takes $V$ to be the tangent 
space $T_q$ at a point $q \in Q$, so that $Z$ is $T_q \times T^*_q$, i.e. the 
tangent space to the space $\Gamma$ of the $q$s and $p$s.

\section{Poisson brackets and Noether's theorem}\label{PoissNoet}
We have seen how a single scalar function $H$ on phase space $\Gamma$ determines 
the evolution of the system via a combination of partial differentiation (the 
gradient of $H$) with the symplectic matrix. We now express these ideas in terms 
of Poisson brackets.

 For our purposes, Poisson brackets will have three main advantages; which will be 
discussed in the following order in the Subsections below. Poisson brackets:\\
\indent (i) give a neat expression for the rate of change of any dynamical 
variable;\\
\indent (ii) give a version of Noether's theorem which is more simple and powerful 
(and even easier to prove!) than the Lagrangian version; and\\
\indent (iii) lead to the generalized Hamiltonian framework mentioned in Section 
\ref{Glimpsegenl}.\\ 
All three advantages arise from the way the Poisson bracket encodes the way that a 
scalar function determines a (certain kind of) vector field.

\subsection{Poisson brackets introduced}\label{pbintro}
The rate of change of any dynamical variable $f$, taken as a scalar function on 
phase space $\Gamma$, $f(q,p) \in \mathR$, is given (with summation convention) by
\be
 \frac{df}{dt} = {\dot q}^i \frac{\pl f}{\pl q^i} + {\dot p}_i \frac{\pl f}{\pl 
p_i} \;.
\label{naivehamdfdt}
\ee   
(If $f$ is time-dependent, $f: (q,p,t) \in \Gamma \times \mathR \mapsto f(q,p,t) 
\in \mathR$, the right-hand-side includes a term $\frac{\pl f}{\pl t}$. But on 
analogy with how our discussion of Lagrangian mechanics imposed scleronomic 
constraints, a time-independent work-function etc., we here set aside the 
time-dependent case.) Applying Hamilton's equations, this is
\be
 \frac{df}{dt} = \frac{\pl H}{\pl p_i} \frac{\pl f}{\pl q^i} - \frac{\pl H}{\pl 
q^i} \frac{\pl f}{\pl p_i} \; .
\label{intropb}
\ee
This suggests that we define the Poisson bracket of any two such functions 
$f(q,p), g(q,p)$ by
\be
\{f, g\} := \frac{\pl f}{\pl q^i} \frac{\pl g}{\pl p_i}  - \frac{\pl f}{\pl p_i} 
\frac{\pl g}{\pl q^i}  \; ;
\label{naivedefinepb}
\ee
so that the rate of change of $f$ is given by 
\be
 \frac{df}{dt} = \{f, H \}  \;.
\label{naivedfdtaspb}
\ee 

In terms of the $2n$ coordinates $\xi^{\al}$ (eq. \ref{definexisupal}) and the 
matrix elements  $\o^{{\al\bb}}$ of $\o$ (eq. \ref{HamxisupalWithOmega}), we can 
write eq. \ref{intropb} as
\be
\frac{df}{dt} = ({\pl}_{\al} f){\dot \xi}^{\al} = ({\pl}_{\al} f) 
\o^{\al\bb}(\pl_{\bb}H) \;\; ;
\label{rocfromxi}
\ee
and so we can define the Poisson bracket by
\be
\{f, g\} := ({\pl}_{\al} f) \o^{\al\bb} (\pl_{\bb}g) \equiv 
\frac{\pl f}{\pl {\xi}^{\al}} \o^{\al\bb} \frac{\pl g}{\pl {\xi}^{\bb}}
 \;\; .
\label{definepbfromxi}
\ee

In matrix notation:  writing the naive gradients of $f$ and of $g$ as column 
vectors $\nabla f$ and $\nabla g$, and 
writing $\; \tilde{} \;$ for  transpose, we have at any point $z = (q,p) \in 
\Gamma$:
\be
\{f, g\} (z) = {\tilde {\nabla f}}(z). \o. \nabla g (z).
\label{pbasmatrix}
\ee

With these definitions of the Poisson bracket, we readily infer the following  
five results. (Later discussion will bring out the significance of some of these; 
in particular, Section \ref{Glimpsegenl} will take some of them  to jointly define 
a primitive Poisson bracket for a generalized Hamiltonian mechanics.)\\
\indent (1): Since the Poisson bracket is antisymmetric, $H$ itself is a constant 
of the motion:
\be
\frac{dH}{dt} = \{H , H \} \equiv 0 \; .
\label{Hcotm}
\ee
\indent (2): The Poisson bracket of a product is given by ``Leibniz's rule'': i.e. 
for any three functions $f,g,h$, we have
\be
\{f, h \cdot g \} = \{f, h \}\cdot g + h \cdot \{f, g \} \; .
\label{Pbofproductfirst}
\ee
\indent (3): Taking the Poisson bracket as itself a dynamical variable, its 
time-derivative is given by a ``Leibniz rule''; i.e. the Poisson bracket behaves 
like a product:
\be
\frac{d}{dt}\{f, g \} = \{\frac{df}{dt}, g \} + \{f, \frac{dg}{dt} \} \; .
\label{Pblikeproduct}
\ee 
\indent (4): The Jacobi identity (easily deduced from (3)):
\be
\{\{f, h \}, g \} + \{\{g, f \}, h \} + \{\{h, g \}, f \}= 0 \;\; .
\label{Jacobiidentityfirst}
\ee
\indent (5): The  Poisson brackets for the $q$s, $p$s and $\xi$s are: 
\begin{eqnarray}
\{\xi^{\al}, \xi^{\bb} \} = \o^{\al\bb} \;\; ; \;\; {\rm{i.e.}} \;\; \\ 
\{q^i, p_j \} = \dd^i_j \;\;, \;\;\;\; \{q^i, q^j \} = \{p_i, p_j \} = 0 \; .
\label{fundlPBs}
\end{eqnarray}

Eq. \ref{fundlPBs} is very important, both for general  theory and for 
problem-solving. The  reason is that preservation of these Poisson brackets, by a 
smooth transformation of the $2n$ variables $(q,p) \rightarrow (Q(q,p), P(q,p))$, 
is necessary and sufficient for the transformation being canonical. Besides, in 
this equivalence `canonical' can be understood both in the usual elementary sense 
of preserving the form of Hamilton's equations, for any Hamiltonian function, and 
in the geometric sense of preserving the symplectic form (explained in (5) of 
Section \ref{formsassoclin}, and for manifolds in Section \ref{geomperspHam}).\\
\indent Note here that, as the phrase `for any Hamiltonian function' brings out, 
the notion of a canonical transformation is independent of the forces on the 
system as encoded in the Hamiltonian. That is: the notion is a matter of 
$\Gamma$'s geometry---as we will emphasise in Section \ref{geomperspHam}.

 But (as I announced in Section \ref{Hammechspreamb}) I will not need to go into 
many details about canonical transformations, essentially because this paper does 
not aim to survey the whole of Hamiltonian mechanics, or even all that can be said 
about reducing problems, e.g. by finding simplifying canonical transformations. It 
aims only to survey the way that symmetries and conserved quantities  effect such 
reductions. In the rest of this Subsection, I begin describing Poisson brackets' 
role in this, in particular Noether's theorem. But the description can only be 
completed once we have the geometric perspective on Hamiltonian mechanics, i.e. in 
Section \ref{noetcomplete}.

\subsection{Hamiltonian vector fields}\label{hamnvfs}
Section \ref{DetsysevolngradH} described how the symplectic matrix enabled the 
scalar function $H$ on $\Gamma$ to determine a vector field $X_H$. 
The previous Subsection showed how the Poisson bracket expressed any dynamical 
variable's rate of change along $X_H$. We now bring these ideas together, and 
generalize.

 Recall that a vector $X$ at a point $x$ of a manifold $M$ can be identified with 
a directional derivative operator at $x$ assigning to each smooth function $f$ 
defined on a neighbourhood of $x$ its directional derivative along any curve that 
has $X$ as its tangent vector. Thus recall the Lagrangian definition of the 
dynamical vector field,  eq. \ref{LagDelta} in Section \ref{geomperspLag}. 
Similarly here: the dynamical vector field $X_H =: D$ is a derivative operator on 
scalar functions, which can be written in terms the Poisson bracket:
\be
D := X_H = \frac{d}{dt} = {\dot q}^i \frac{\pl }{\pl q^i} + {\dot p}_i \frac{\pl 
}{\pl p_i} = 
\frac{\pl H}{\pl p_i} \frac{\pl }{\pl q^i} - \frac{\pl H}{\pl q^i} \frac{\pl }{\pl 
p_i} = \{\cdot, H\}
\;.
\label{HamDelta0}
\ee  

But this point applies to any smooth scalar, $f$ say, on $\Gamma$. That is:
although we think of $H$ as the energy that determines the real physical 
evolution, the mathematics is of course the same for such an $f$. So any such 
function determines a vector field, $X_f$ say, on $\Gamma$ that generates what the 
evolution ``would be if $f$ was the Hamiltonian''. Thinking of the integral curves 
as parametrized by $s$, we have
\be
X_f = \frac{d}{ds} = \{\cdot, f \} \; .
\label{defineX_fusual}
\ee 
$X_f$ is called the {\em Hamiltonian vector field} of (for) $f$; just as, for the 
physical Hamiltonian, $f \equiv H$, Section \ref{DetsysevolngradH} called $X_H$ 
`{\em the} Hamiltonian vector field'.

The notion of a Hamiltonian vector field will be crucial for what follows, not 
least for Noether's theorem in the very next Subsection. For the moment, we just 
make two remarks which we will need later. 

So every scalar $f$ determines a Hamiltonian vector field $X_f$. But note that the 
converse is false:  not every vector field $X$ on $\Gamma$ is the Hamiltonian 
vector field of some scalar. For a vector field (equations of motion) $X$, with 
components $X^{\al}$ in the coordinates  $\xi^{\al}$ defined by eq. 
\ref{definexisupal}
\be
{\dot \xi}^{\al} = X^{\al}(\xi) \;\; ,
\ee
there need be no scalar $H: \Gamma \rightarrow \mathR$ such that, as required by 
eq. \ref{HamxisupalWithOmega},
\be
X^{\al} = \o^{\al\bb}\pl_{\bb}H \; .
\ee
This is the same point as in (ii) of Section \ref{Legtrsfmnvarnal}: that 
Hamilton's equations have the special feature that all the right hand sides  are, 
up to a sign, partial derivatives of a {\em single} function $H$---a feature that 
underpins the possibility of expressing the equations of motion by variational 
principles.

We also need to note under what condition is a vector field $X$ Hamiltonian; (this 
will bear on Noether's theorem). The answer is: $X$ is locally Hamiltonian, i.e. 
there is locally a scalar $f$ such that $X = X_f$, iff $X$ generates a 
one-parameter family of canonical transformations. We will give a modern geometric 
proof of this in Section \ref{noetcomplete}. For the moment, we only need to note, 
as at the end of Section \ref{pbintro}, that here `canonical  
transformation' can be understood in the usual elementary sense as a 
transformation of  $\Gamma$ that preserves the form of Hamilton's equations (for 
any Hamiltonian); or equivalently, as preserving the Poisson bracket; or 
equivalently, as preserving the  symplectic form (to be defined for manifolds, in 
Section \ref{geomperspHam}).

\subsection{Noether's theorem}\label{noetHamsimple}

\subsubsection{An apparent ``one-liner'', and three claims}\label{443A} 
In the Hamiltonian framework, the core of the proof of Noether's theorem is very 
simple; as follows.  
The Poisson bracket is obviously antisymmetric. So for any scalar functions $f$ 
and $H$, we have 
\be
X_f(H) \equiv \frac{dH}{ds} \equiv \{H, f\} = 0 \;\;\;\; {\mbox{ iff }} \;\; \;\;
0 = \{f,H\} = X_H(f) \equiv D(f) \; \; .
\label{naivenoetham} 
\ee
In words: $H$ is constant under the flow of the vector field $X_f$ (i.e. under 
what the evolution would be if $f$ was the Hamiltonian) iff $f$ is constant under 
the dynamical flow $X_H \equiv D$.

This ``one-liner'' is the Hamiltonian version of Noether's theorem! There are 
three claims here. The first two relate back to the Lagrangian  version of the 
theorem. The third is about the definition of a (continuous) symmetry for a 
Hamiltonian system, and so about how we should formulate the Hamiltonian version 
of Noether's theorem. I will state all three claims, but in this Subsection 
justify only the first two. For it will be convenient to postpone the third till 
after we have introduced some modern geometry (Section \ref{noetcomplete}).   

 First, for eq. \ref{naivenoetham} to deserve the name `Noether's theorem', I need 
to show that  it encompasses Section \ref{NoetherLag}'s Lagrangian version of 
Noether's theorem (despite the trivial proof!).\\
\indent Second, in order to justify my claim that the Hamiltonian version of 
Noether's theorem is more powerful than the Lagrangian version, I need to show 
that eq. \ref{naivenoetham} says more than that version, i.e. that it covers more 
symmetries.

\indent To state the third claim, note first that we expect a Hamiltonian version 
of Noether's theorem to say something like: {\em to every continuous symmetry of a 
Hamiltonian system, there corresponds a conserved quantity}. Here, we expect a 
`continuous symmetry' to be defined by a vector field on $\Gamma$ (or by its 
flow). Indeed, a {\em symmetry} of a Hamiltonian system is usually defined as a 
transformation of $\Gamma$ that:\\
\indent (1) is canonical; (a condition independent of the forces on the system as 
encoded in the Hamiltonian: a matter of $\Gamma$'s intrinsic geometry); and also\\
\indent (2) preserves the Hamiltonian function; (a condition obviously dependent 
on the Hamiltonian).\\
Accordingly, a {\em continuous symmetry} is defined as a vector field on $\Gamma$ 
that generates  a one-parameter family of such transformations; (or as such a 
field's flow, i.e. as the family itself).

But with this definition of `continuous symmetry' (of a Hamiltonian system), eq. 
\ref{naivenoetham} seems to suffer from two {\em lacunae}, if taken to express 
Noether's theorem, that to every continuous symmetry there corresponds a conserved 
quantity. Agreed, the rightward implication of eq. \ref{naivenoetham} provides, 
for a vector field $X_f$ with property (2), the conserved quantity $f$. But there 
seem to be two {\em lacunae}:\\
\indent (a): eq. \ref{naivenoetham} is silent about whether $X_f$ has property 
(1), i.e. generates canonical transformations.\\
\indent (b): eq. \ref{naivenoetham} considers only Hamiltonian vector fields, i.e. 
vector fields $X$ induced by some $f$, $X = X_f$. But as noted at the end of 
Section \ref{hamnvfs}, there are countless vector fields on $\Gamma$ that are not 
Hamiltonian. If such a field could be a continuous symmetry, eq. 
\ref{naivenoetham}'s rightward implication would fall short of saying that to {\em 
every} continuous symmetry, there corresponds a conserved quantity.

So the third claim I need is that these {\em lacunae} are illusory. In fact, a 
single result will deal with both (a) and (b). Namely, it  will suffice to show 
that a vector field $X$ on $\Gamma$ has property (1), i.e. generates canonical 
transformations, iff it is Hamiltonian, i.e. induced by some $f$, $X = X_f$. But I 
postpone showing this till we have more modern geometry in hand; cf. Section 
\ref{noetcomplete}.

\subsubsection{The relation to the Lagrangian version}\label{443B} 
On the other hand, we can establish the first two claims with the elementary 
apparatus so far developed. I will concentrate on justifying the first claim; that 
will also make the second claim clear.

For the first claim, we need to show that:\\
\indent (i): to any variational symmetry of the Lagrangian $L$, i.e. a vector 
field $X$ on $Q$ obeying eq. \ref{definevarnlsymy2}, there corresponds a vector 
field $X_f$ on $\Gamma$ for which $X_f(H) = 0$; and \\
\indent (ii): the correspondence in (i) is such that the scalar $f$ can be taken 
to be (the Hamiltonian version of) the momentum $p_X$ conjugate to $X$, defined by 
eq. \ref{defconjugmommX} (or geometrically, by \ref{pXascontraction}).

It will be clearest to proceed in two stages.\\
\indent  (A): First,  I will show (i) and (ii).\\
\indent  (B): Then I will discuss how (A) relates to the usual definition  of a 
symmetry of a Hamiltonian system.

(A): The easiest way to show (i) and (ii) is to use the fact discussed after eq. 
\ref{pXispn!}, that every variational  symmetry $X$  arises, around a point where 
it is non-zero, from a cyclic coordinate in some local system of coordinates. 
(Recall that this follows from the basic ``rectification'' theorem securing the 
local existence and uniqueness of solutions of ordinary differential equations.) 
That is, there is some coordinate system $(q)$ on some open subset of $X$'s domain 
of definition on $Q$ such that \\
\indent (a): $X$ being a variational symmetry is equivalent to $q^n$ being cyclic, 
i.e. $\frac{\pl L}{\pl q^n} = 0$;\\
\indent (b): the  momentum $p_X$, which the Lagrangian theorem says is conserved,  
is the elementary generalized momentum $p_n := \frac{\pl L}{\pl {\dot q}^n}$.

So suppose given a variational symmetry $X$, and a coordinate system $(q)$ 
satisfying (a)-(b). Now we recall that the Legendre transformation, i.e. the 
transition between Lagrangian and Hamiltonian frameworks, does not  ``involve the 
dependence on the $q$s''.  More precisely, we recall eq. \ref{Hqn=Lqn}, $\frac{\pl 
H}{\pl q^n} \; = \; - \frac{\pl L}{\pl q^n}$. Now consider $p_n: \Gamma 
\rightarrow \mathR$. This $p_n$ will do as  the function  $f$ required in (i) and 
(ii) above, since
\be
X_{p_n}(H) \equiv \{H, p_n \} = \frac{\pl H}{\pl q^n} = - \frac{\pl L}{\pl q^n} = 
0.
\label{XpnH}
\ee
Applying eq. \ref{naivenoetham} to eq. \ref{XpnH}, we deduce that $p_n$, i.e. the 
$p_X$ of the Lagrangian theorem, is conserved.

(Hence my remark after eq. \ref{Hqn=Lqn}, that the elementary result that $p_n$ is 
conserved iff $q^n$ is cyclic, underpins the Hamiltonian version of Noether's 
theorem;  
just as the corresponding  Lagrangian result underpins the Lagrangian version of 
Noether's theorem: cf. discussion after eq. \ref{pXispn!}.)

(B): I agree that this simple proof seems {\em suspiciously} simple. Besides, the 
suspicion grows when you notice that my argument in (A) has not used a definition 
of a symmetry, in particular a continuous symmetry, of a Hamiltonian system 
(contrast Section \ref{VecfieldsSymmies}). As discussed in Section \ref{443A}, we 
expect  a Hamiltonian version of Noether's theorem to say `to every continuous 
symmetry of a Hamiltonian system there corresponds a conserved quantity'; where a 
continuous symmetry is a vector field that (1) generates canonical transformations 
and (2) preserves the Hamiltonian. So the argument in (A) is suspicious since, 
although eq. \ref{XpnH}, or the left hand side of eq. \ref{naivenoetham}, 
obviously expresses property (2), i.e. preserving the Hamiltonian, the argument in 
(A) seems to nowhere use property (1), i.e. the symmetry generating  canonical 
transformations. 

But in fact, all is well. The reason why lies in the fact mentioned in (i), (a) of 
Section \ref{Hammechspreamb}: that  every point  transformation (together with its 
lift to $TQ$) defines a corresponding canonical transformation on $T^*Q$. That is 
to say: property (1) is secured by the fact that the Lagrangian Noether's theorem 
of Section \ref{NoetherLag} is restricted to symmetries induced by point 
transformations. \\
\indent In other words, in terms of the vector field (variational symmetry) $X$ 
given us by (a) in (A) above: one can check that $X$ defines a vector field on 
$\Gamma$ (equivalently: a one-parameter family of transformations on $\Gamma$) 
that is canonical, i.e. preserves Hamilton's equations or equivalently the 
symplectic form. Indeed, one can easily check that, once we rectify the Lagrangian 
variational symmetry $X$, so that it generates the rectified one-parameter family 
of point transformations: $q_i = {\rm {const}}, i \neq n; q_n \mapsto q_n + 
\epsilon$, the vector field that $X$ defines on $\Gamma$  is precisely the field  
$X_{p_n}$ chosen above.\footnote{Details about point transformations on $Q$ 
defining a canonical transformation on $T^*Q$, and lifting the vector field $X$ to 
$\Gamma$, can be found: (i) using traditional terms, in Goldstein et al. (2002: 
375-376) and Lanczos (1986: Chapter VII.2); (ii) using modern geometric terms (as 
developed in Section \ref{geomperspHam}), in Abraham and Marsden (1978: Sections 
3.2.10-3.2.12) and Marsden and Ratiu (1999: Sections 6.3-6.4).}
  
Finally, the discussion in (B) also vindicates the second claim in Section 
\ref{443A}: that the Hamiltonian version of Noether's theorem, eq. 
\ref{naivenoetham}, {\em says more} than the Lagrangian version, i.e. covers more 
symmetries. This follows from the fact (announced in (i) (b) of Section 
\ref{Hammechspreamb}) that there are canonical transformations {\em not} induced 
by a point transformation (together with its lift).\\
\indent In elementary discussions, this is often expressed in terms of canonical 
transformations being allowed to ``mix'' the $q$s and $p$s. But a more precise, 
and geometric, statement is the result announced at the end of Section 
\ref{hamnvfs} (whose proof is postponed to Section \ref{noetcomplete}): that the 
condition for a vector field on $\Gamma$ to generate a one-parameter family of 
canonical transformations is merely that it be a Hamiltonian vector field. That 
is: for {\em any} scalar $f: \Gamma \rightarrow \mathR$, the vector field $X_f$ 
generates such a family.

In this sense, canonical transformations are two a penny (also known as: a dime a 
dozen!). So it is little wonder that most discussions emphasise the {\em other} 
condition, i.e. property (2):  that $X_f$ preserve the Hamiltonian, $X_f(H) = 0$. 
Only very special $f$s will satisfy $X_f(H) = 0$; and if we are given $H$ (in 
certain coordinates $q,p$), it can be very hard to find (the coordinate expression 
of) such an $f$.\\
\indent Indeed, when Jacobi first propounded the theory of canonical 
transformations, in his {\em Lectures on Dynamics} (1842), he was of course aware 
of this. Accordingly, he pointed out that in theoretical mechanics, it was often 
more fruitful to first consider an $f$ (equivalently: a canonical transformation), 
and then cast about for a Hamiltonian that it preserved.  He wrote: `The main 
difficulty in integrating a given differential equation lies in introducing 
convenient variables, which there is no rule for finding. Therefore we must travel 
the reverse path and after finding some notable substitution, look for problems to 
which it can be successfully applied'; (quoted in Arnold (1989, p. 266)). The fact 
that  Jacobi solved many previously intractable problems bears witness to the 
power of this strategy, and of his theory of canonical transformations.

We can sum up this Subsection in two comments:---\\
\indent (1) In Hamiltonian mechanics, Noether's theorem is a biconditional, an 
`iff' statement. Not only does a Hamiltonian symmetry---i.e. a vector field $X$ on 
$\Gamma$ that generates canonical transformations (equivalently: preserves the 
symplectic form, or the Poisson bracket) and preserves the Hamiltonian, $X(H) = 
0$---provide a constant of the motion. Also, given a constant of the motion $f: 
\Gamma \rightarrow \mathR$, there is a symmetry of the Hamiltonian, viz. the 
vector field $X_f$. (Or if one prefers the integral notion of symmetry: the flow 
of $X_f$). This converse implication, from constant to symmetry, contrasts with 
the Lagrangian framework; cf. the end of Section \ref{NoetherLagGeomSubsub}.\\
\indent (2) In elementary Hamiltonian mechanics, Noether's theorem has a very 
simple one-line proof, viz. eq. \ref{naivenoetham}. 

Later, we will return to Noether's theorem. Section \ref{noetcomplete} will 
justify the third claim of Section \ref{443A}, by showing that a vector field 
generates a one-parameter family of canonical transformations iff it is a 
Hamiltonian vector field.  Meanwhile, we end Section \ref{PoissNoet} with a 
comment about  ``iterating'' Noether's theorem, and the distinction  between such 
an iteration and the idea of complete integrability.  

\subsection{Glimpsing the ``complete solution''}\label{Glimpseintby}
Suppose we ``iterate'' Noether's theorem. That is: suppose there are several 
(continuous) symmetries of the Hamiltonian and so several constants of the motion. 
Each will confine the system's time-evolution to a ($2n-1$)-dimensional 
hypersurface of $\Gamma$. In general, the intersection of $k$ such surfaces will 
be a hypersurface of dimension $2n - k$ (i.e. of co-dimension $k$); to which the 
motion is therefore confined. The theory of symplectic reduction (Butterfield 
2006) describes how to do a ``quotiented dynamics'' in this general situation. 
Here, I just remark on one aspect; which will {\em not} be developed in the 
sequel.

{\em Locally}, the rectification theorem secures, for any system, not just several 
constants of the motion, but ``all you could ask for''. Applying the theorem (eq. 
\ref{arnold10de} and \ref{arnold10deparallelized}) to the Hamiltonian vector field 
$X_H$ on $\Gamma$, we infer that locally there are coordinates $\xi^{\al}$ (maybe 
very hard to find!) in which $X_H$ has $2n -1$ components that vanish throughout 
the neighbourhood, while the other component is 1:
\be
X^{\al}_H = 0 {\rm{\;\; for \;\;}} \al = 1,2, \dots, 2n -1 \;\; ; \;\; X^{2n}_H = 
1 \;\; .
\label{XHrectified}
\ee 
So the coordinates $\xi^{\al}, \al = 1,...,2n - 1,$ form $2n - 1$ constants of the 
motion. They are functionally independent, and all other constants of the motion 
are functions of them; (cf. point (ii) after eq. \ref{arnold10deparallelized}). So 
the motion is confined to the one-dimensional intersection of the $2n - 1$ 
hypersurfaces, each of co-dimension 1. That is to say, it is confined  to the 
curve given by: $\xi^{\al} = {\rm{const}}, \al = 1,...,2n-1, \xi^{2n} = t$.

To this, Noether's theorem eq. \ref{naivenoetham} adds the physical idea that each 
such constant of the motion defines a vector field $X_{\xi^{\al}}$ that generates 
a symmetry of the Hamiltonian: 
\be
X_{\xi^{\al}}(H) = 0, {\rm{\;\; for \;\;}} \al = 1,2, \dots, 2n -1 \; .
\ee
In this local sense, the ``complete solution'' of any Hamiltonian system lies in 
the local constants of the motion, or equivalently the local symmetries of its 
Hamiltonian $H$.\\
\indent  To sum up: locally, any Hamiltonian  system is ``completely integrable''. 
But the scare-quotes here are  a reminder that these phrases are usually used with 
other, stronger, meanings: either that there are $2n-1$ {\em global} constants of 
the motion or that the system is completely integrable in the sense of Liouville's 
theorem.

\section{A geometrical perspective}\label{geomperspHam}
In this final Section, we develop the modern geometric description of Hamiltonian 
mechanics. We will build especially on Sections \ref{sympformintro}; one main aim 
will of course be to  complete the discussion of Noether's theorem, begun in 
Section \ref{noetHamsimple}.
 
There will be eight Subsections. First, we introduce the cotangent bundle $T^*Q$. 
Then we collect what we will need about forms. Then we can show that any cotangent 
bundle is a symplectic manifold. This enables us to formulate Hamilton's equations  
geometrically; and to complete the discussion of Noether's theorem. Then we report 
Darboux's theorem, and its relation to reduction of problems. Then we  return to 
the Lagrangian framework, by sketching the geometric formulation of the Legendre 
transformation. Finally, we ``glimpse the landscape ahead'' by mentioning the more 
general  framework for Hamiltonian mechanics that uses Poisson manifolds.

\subsection{Canonical momenta are one-forms: $\Gamma$ as 
$T^*Q$}\label{canlmoma1forms}
So far we have treated the phase  space $\Gamma$ informally: saying just that it 
is a $2n$-dimensional space coordinatized by the $q$s, a smooth coordinate system 
on the configuration manifold  $Q$, and the $p$s, which are canonical momenta 
$\frac{\pl L}{\pl {\dot q}^i}$. 
But we also saw in (3) of Section \ref{tgtble} that at each point $q \in Q$, the 
$p_i$ transform as a 1-form (eq. \ref{psareoneform}). Accordingly we now take the 
physical state of the system to be a point in the cotangent bundle $T^*Q$, the 
$2n$-dimensional manifold whose points are pairs $(q, p)$ with $q \in Q, p \in 
T^*_q$.

\indent I stress that from now on, the symbol $p$ has a (fruitful!) ambiguity, 
between  ``dynamics'' and ``kinematics/geometry''. For $p$ represents both:\\
\indent (A) the conjugate momentum $\frac{\pl L}{\pl {\dot q}}$, which of course 
depends on the choice of $L$; and\\
\indent (B) a point in a fibre $T^*_q$ of the cotangent bundle $T^*Q$ (i.e. a 
1-form or covector); or relatedly: the components $p_i$ of such a 1-form: notions 
that are independent of any choice of a Lagrangian or Hamiltonian.\\
\indent In more detail:---

(A): Recall that in the Lagrangian framework, the basic equations (eq. 
\ref{eqn;lag}, or Newton's second law!) being second-order in time prompts us to 
take the initial $q$ and ${\dot q}$ as chosen independently, with $L$ (encoding 
the forces on the system) then determining the evolution (the Lagrangian  
dynamical vector field $D$)---and so also determining the actual ``realized'' 
value of ${\dot q}$ at other times as a function of $q$, and so ultimately, of 
$t$. Similarly here: Newton's second law being second-order in time prompts us to 
take the initial $q$ and $p$ as  independent, with $H$ (encoding the forces on the 
system) then determining the evolution (the Hamiltonian dynamical vector field 
$D$)---and so also determining the actual value of $p$ at other times  as a 
function of $q$, and so ultimately, of $t$. Besides,  by passing via the Legendre 
transformation  back to the Lagrangian framework, one can check that the later 
actual value of $p$ is determined to equal $\frac{\pl L}{\pl {\dot q}}$.

(B): But $p$ also represents any 1-form (so that $p_i$ represents the 1-form's 
coordinates). Here, we need to recall three points:---

\indent (i): A local coordinate system (a chart) on $Q$ defines a  basis in the 
tangent space $T_q$ at any point $q$ in the chart's domain. As usual, I write the 
chart's coordinate functions as $q^i$. So I shall temporarily denote the chart by 
$[q]$, so that there are coordinate functions $q^i: {\rm {dom}}([q]) \rightarrow 
\mathR$. I write elements of the coordinate basis as usual, as $\frac{\pl}{\pl 
q^i}$.\\
\indent (ii): The chart $[q]$ thereby also defines a dual basis $dq^i$ in the 
cotangent space $T^*_q$ at any $q \in {\rm {dom}}([q])$.\\
\indent\indent (Here I recall, {\em en passant}, that the isomorphism at each $q$ 
between $T_q$ and $T^*_q$, that maps the basis element $\frac{\pl}{\pl q^i} \in 
T_q$ to the one-form  $dq^i$ in the  dual basis, is basis-{\em dependent}. A 
different basis $\frac{\pl}{\pl q'^i}$ would give a different isomorphism. Cf. the 
discussion in (1) of Section \ref{formsassoclin}.)\\
\indent (iii): Putting (i) and (ii) together: the chart $[q]$ thereby also induces 
a local coordinate system on a neighbourhood of the cotangent bundle around any 
point $(q, p) \in T^*Q$ with $q \in {\rm dom}([q])$ and $p \in T^*_q$.

Putting (i)-(iii) together: the coordinates of any  point $(q,p)$ in $T^*Q$ in 
such a coordinate system are usually also written as $(q,p)$. That is: $p$ is used 
for the components of {\em any} 1-form, in the basis $dq^i$ dual to a coordinate 
basis $\frac{\pl }{\pl q^i}$. So, similarly to (i) above: I will write this 
induced chart on $T^*Q$ as $[q,p]$.

(C): Taken together, points  (A) and (B) prompt a question:
\begin{quote} 
Why should an evolution from an arbitrary initial state $\in T^*Q$ have the 
property that:---\\
{\em if} we choose to express\\
\indent (i) its configuration, $q_0$ say,  in terms of an arbitrary initial 
coordinate system $[q]$ on $Q$, and\\
\indent (ii) its momenta $\frac{\pl L}{\pl {\dot q}}$ in terms of the basis $dq$ 
dual to the coordinate  basis $\frac{\pl }{\pl q}$ at $q_0$:---\\
{\em then}\\
 the  states at a {\em later} time $t$ have {\em their} momenta---which the 
Lagrangian framework tells us must be  $\frac{\pl L}{\pl {\dot q}}$  (cf. 
(A))---equal to their components in the dual basis to the {\em later} coordinate 
basis, i.e. the coordinate basis $\frac{\pl }{\pl q}$ at the later configuration  
$q_t$?

In short: why should the state's components in the dual basis of any coordinate 
basis continue to be equal, as dynamical evolution goes on, to the values of 
canonical momenta i.e. $\frac{\pl L}{\pl {\dot q}}$?    
\end{quote}
A good question. The short answer lies in combining Hamilton's equations for the 
time-derivative of the $p_i$ (eq. \ref{HamFromLagSimple}) with Lagrange's 
equations, and with the fact that the partial derivatives with respect to $q^i$ of 
the Hamiltonian and Lagrangian, $H$ and $L$, are negatives of each other (eq. 
\ref{Hqn=Lqn}). Thus we have:
\be
{\dot p}_i = - \frac{\pl H}{\pl q^i} = \frac{\pl L}{\pl q^i} = 
\frac{d}{dt}\left(\frac{\pl L}{\pl {\dot q}^i}\right) \;\; .
\label{answergoodqn}
\ee
From this it is clear that for any coordinate system, if at $t_0$, $p_i$ is chosen 
to equal $\frac{\pl L}{\pl {\dot q}^i}$, then this will be so at later times. For 
eq. \ref{answergoodqn} forces their time-derivatives to be equal---and so also, 
their later values must be equal. 

So much for the short answer. We will also get more insight into the relations   
between the Lagrangian and Hamiltonian  frameworks in\\
\indent (i) the fact, expounded in Section \ref{cotgtblesymp} below, that any 
cotangent bundle has a natural symplectic structure, independent of the 
specification of any Lagrangian or Hamiltonian function; and\\
\indent (ii) some further details about the Legendre transformation, which is 
further discussed in Section \ref{geomlegetrsfmn}.

\subsection{Forms, wedge-products and exterior derivatives}\label{introduceforms}
As I said at the end of Section \ref{interpareas},  this paper can largely avoid 
the theory of forms. For what follows (especially Section \ref{noetcomplete}), I 
need to recall only:\\
\indent (i) the idea of forms of various degrees, together comprising the exterior 
algebra, and equipped with operations of wedge-product and contraction (Section 
\ref{452A});\\
\indent (ii) the ideas of differential forms, the exterior derivative, and of 
exact and closed forms (Section \ref{452B}).

\subsubsection{The exterior algebra; wedge-products and contractions}\label{452A} 
We begin by recalling some ideas of Sections \ref{interpareas} and 
\ref{formsassoclin}. Let us again begin with the simplest possible case, 
$\mathR^2$, considered as a vector space: not as a manifold with a copy of itself 
as tangent space at each point. \\
\indent If $\al, \bb$ are covectors, i.e. elements  of $(\mathR^2)^*$, we define 
their {\em wedge-product}, an antisymmetric bilinear form on $\mathR^2$, by
\be
\al \wedge \bb: (v,w) \in \mathR^2 \times \mathR^2 \mapsto (\al (v))(\bb (w)) - 
(\al (w))(\bb (v)) \in \mathR \;\; .
\ee
Let us write the standard basis elements of $\mathR^2$ as $\frac{\pl}{\pl q}$ and 
$\frac{\pl}{\pl p}$, with elements of $\mathR^2$ having components $(q,p)$ in this 
basis; and let us write the elements of the dual basis  as $dq, dp$. Recalling the 
definition of the area form $A$, eq. \ref{defineAR2}, we deduce that 
$A$ is $dq \wedge dp$.

\indent Similarly for $\mathR^{2n}$. Recall that the symplectic matrix  defines an 
antisymmetric bilinear form on $\mathR^{2n}$ by eq. \ref{explainARn}. The value on 
a pair $(q,p) \equiv (q^1,...q^n; p_1,...,p_n), (q',p') \equiv (q'^1,...q'^n; 
p'_1,...,p'_n)$ is the sum of the signed areas of the $n$ parallelograms formed by 
the projections of the vectors $(q,p), (q',p')$ onto the $n$ pairs of coordinate 
planes. This is a sum of $n$ wedge-products. That is to say: if we write the 
standard basis elements as $\frac{\pl}{\pl q^i}$ and $\frac{\pl}{\pl p_i}$, this 
form is  $\omega := \Sigma_i \; dq^i \wedge dp_i$. It has the action on $\mathR^n 
\times \mathR^n$:
\be
(q^i\frac{\pl}{\pl q^i} + p_{i}\frac{\pl}{\pl p_i}, q'^i\frac{\pl}{\pl q^i} + 
p'_{i}\frac{\pl}{\pl p_i}) \mapsto 
\Sigma^n_{i=1} \; q^i p'_{i} - q'^{i} p_i \;\;.
\label{explainARnagain}
\ee

In general, if $V,W$ are two (real finite-dimensional) vector spaces,  we define: 
$L(V,W)$ to be the vector space of linear maps from $V$ to $W$; $L^k(V,W)$ to be 
the vector space of $k$-multilinear maps from $V \times V \times .... \times V$ 
($k$ copies) to $W$; and $L^k_a(V,W)$ to be the subspace of $L^k(V,W)$ consisting 
of (wholly) antisymmetric maps.\\
\indent We then define $\O^k(V) := L^k_a(V,\mathR)$ for $k = 1,2, 
...,{\rm{dim}}(V)$, so that $\O^1(V) = V^*$. We also set $\O^0(V) := \mathR$. 
$\O^k(V)$ is called the space of {\em (exterior) $k$-forms} on $V$. If dim$(V) = 
n$, then dim$(\O^k(V)) = \left(\begin{array}{c}
n\\
k \\
\end{array}\right)$.

\indent The wedge-product, as defined above, can be extended to be an operation 
that defines, for $\al \in \O^k(V), \bb \in \O^l(V)$, an element $\al \wedge \bb 
\in \O^{k+l}(V)$. We can skip the details: suffice it to say that the idea is to 
take tensor products as in (3) of Section \ref{formsassoclin}, and 
anti-symmetrize. 

But to complete our discussion of Noether's theorem (in Section 
\ref{noetcomplete}), we will need the definition of the {\em contraction}, (also 
known as: {\em interior product}), of a $k$-form $\al \in \O^k(V)$ with a vector 
$v \in V$. We shall write this as  ${\bf i}_v \al$. (It is also written with a 
hook notation.) We define the contraction ${\bf i}_v \al$ to be the $(k-1)$-form 
given by:
\be
{\bf i}_v \al (v_2,...,v_k) := \al(v,v_2,...,v_k) \; .
\label{definecontraction}
\ee
It follows, for example, that contraction distributes over the wedge-product {\em 
modulo} a sign, in the following sense. If $\al$ is a $k$-form, and $\bb$ a 
1-form, then
\be
{\bf i}_v (\al \wedge \bb) = ({\bf i}_v \al) \wedge \bb + (-1)^k \al \wedge ({\bf 
i}_v \bb) \; .
\label{contrdistribwedge}
\ee

The direct sum  of the vector  spaces $\O^k(V), k = 0,1,2,...,{\rm{dim}}(V) =: n$, 
has dimension $2^n$. When this direct sum is considered as equipped with the 
wedge-product $\wedge$ and contraction ${\bf i}$, it is called the {\em exterior 
algebra} of $V$, written $\O(V)$.

\subsubsection{Differential forms; the exterior derivative; the Poincar\'{e} 
Lemma}\label{452B}
 We extend the discussion given in Section \ref{452A} to a manifold $M$ of 
dimension $n$, taking all the tangent spaces $T_x$ at $x \in M$  as copies of the 
vector space $V$, and requiring fields of forms to be suitably smooth.

We begin by saying that a (smooth) scalar function $f: M \rightarrow \mathR$ is a 
0-form field. Its {\em differential} or {\em gradient}, $df$, as defined by its 
action on all vector fields $X$, viz. mapping them to $f$'s directional derivative 
along $X$
\be
df(X) := X(f) 
\ee
 is a 1-form (covector) field, called a {\em differential 1-form}. \\
\indent The set ${\cal F}(M)$ of all smooth scalar functions forms an 
(infinite-dimensional) vector space, indeed a ring, under pointwise operations. We 
write the set of vector fields on $M$ as ${\cal X}(M)$, or as ${\cal T}^1_0(M)$; 
and the set of covector fields, i.e. differential 1-forms, on $M$ as ${\cal 
X}^*(M)$, or as ${\cal T}^0_1(M)$.  (So superscripts indicate the contravariant 
order, and  subscripts the covariant order.)\\
\indent Accordingly, we define: $\O^0(M) := {\cal F}(M)$; $\O^1(M) = {\cal 
T}^0_1(M)$; and so on. In short: $\O^k(M)$ is the set of smooth fields of exterior 
$k$-forms on the tangent  spaces of $M$.

The wedge-product, as defined in Section \ref{452A}, can be extended to the 
various $\O^k(M)$. We form the direct sum  of the (infinite-dimensional) vector  
spaces $\O^k(M), k = 0,1,2,...,{\rm{dim}}(V) =: n$, and consider it as equipped 
with this extended wedge-product. We call it the {\em algebra of exterior 
differential forms} on $M$, written $\O(M)$.

Similarly, contraction, as defined in Section \ref{452A}, can be extended to 
$\O(M)$.  On analogy with eq. \ref{definecontraction}, we define, for $\al$ a 
$k$-form field on $M$, and $X$ a vector field on $M$, the contraction ${\bf i}_X 
\al$ to be the $(k-1)$-form given, at each point $x \in M$, by:
\be
{\bf i}_X \al (x): (v_2,...,v_k) \mapsto \al(x)(X(x),v_2,...,v_k) \in \mathR \; .
\label{definecontractionfields}
\ee

\indent The {\em exterior derivative} is a  differential operator on $\O(M)$ that 
maps a $k$-form field to a $(k+1)$-form field. In particular, it maps a scalar $f$ 
to its differential (gradient) $df$. Indeed, it is the {\em unique} map from the 
$k$-form fields to the $(k+1)$-form fields ($k = 1,2,...,n$) that generalizes the 
elementary notion of gradient $f \mapsto df$, subject to certain natural 
conditions.

 To be precise: one can show that there is a unique family of maps $d^k: \O^k(M) 
\rightarrow \O^{k+1}(M)$, all of which, for simplicity, we write as $\bf d$, such 
that:\\
\indent (a): If $f \in {\cal F}(M)$, ${\bf d}(f) = df$.\\
\indent (b): $\bf d$ is $\mathR$-linear; and distributes across the wedge-product, 
{\em modulo} a sign. That is: for $\al \in \O^k(M), \bb \in \O^l(M)$, ${\bf d}(\al 
\wedge \bb) = ({\bf d}\al) \wedge \bb + (-1)^k \al \wedge ({\bf d}\bb).$ (Cf. eq. 
\ref{contrdistribwedge}.)\\
\indent (c): ${\bf d}^2 := {\bf d} \circ {\bf d} \equiv 0$; i.e. for all $\al \in 
\O^k(M)$ $d^{k+1} \circ d^k(\al) \equiv 0$. (This condition looks strong, but is 
in fact natural. For its motivation, it must here suffice to say that it 
generalizes the fact in elementary vector calculus, that the curl of any gradient 
is zero: $\nabla \wedge (\nabla f) \equiv 0$.)\\
\indent (d): ${\bf d}$ is a {\em local operator}; i.e. for any $x \in M$ and any 
$k$-form $\al$, ${\bf d}\al(x)$ depends only on $\al$'s restriction to any open 
neighbourhood of $x$; more precisely, we define for any open set $U$ of $M$, the 
vector space $\O^k(U)$ of $k$-form fields on $U$, and  then require that
\be
{\bf d}(\al \mid_U) = ({\bf d} \al)\mid_U  \; \; .
\label{extderivlocal}
\ee

To express ${\bf d}$ in terms of coordinates:  if $\al \in \O^k(M)$, i.e. $\al$ is 
a $k$-form on $M$, given in coordinates by
\be
\al = \al_{i_1 \dots  i_k} \; dx^{i_1} \wedge \cdots \wedge dx^{i_k} \;\;\; 
({\rm{sum \; on}} \;\;  i_1 < i_2 < \dots < i_k),
\label{kformincooords}
\ee 
then one proves that the exterior derivative is
\be
{\bf d}\al \;\; = \;\; \frac{\pl \al_{i_1 \dots  i_k}}{\pl x^j} \; dx^j \wedge 
dx^{i_1} \wedge \cdots \wedge dx^{i_k}\;\;\; ({\rm{sum \; on \; all}} \; j \; {\rm 
{and}} \;  i_1 , \dots < i_k),
\label{extderivkformincooords}
\ee 

We define $\al \in \O^k(M)$ to be:\\
\indent {\em exact} if there is a $\bb \in \O^{k-1}(M)$ such that $\al = {\bf 
d}\bb$; (cf. the elementary definition of an exact differential);\\
\indent {\em  closed} if ${\bf d}\al = 0$.

It is immediate from condition (c) above, ${\bf d}^2 = 0$, that  every exact form 
is closed. The converse is ``locally true''. This important result is the {\em 
Poincar\'{e} Lemma}; (and we will use it in Section \ref{noetcomplete}'s  closing 
discussion of Noether's theorem).\\
\indent  To be precise:  for any open set $U$ of $M$, we define (as in condition 
(d) above) the vector space $\O^k(U)$ of $k$-form fields on $U$. Then the {\em 
Poincar\'{e} Lemma} states that if $\al \in \O^k(M)$ is closed, then at every $x 
\in M$ there is a neighbourhood $U$ such that $\al\mid_U \; \in \; \O^k(U)$ is 
exact.

We will also need (again, for Section \ref{noetcomplete}'s  discussion of 
Noether's theorem) a useful formula relating the Lie derivative, contraction and 
the exterior derivative. Namely: {\em Cartan's magic formula}, which says that if 
$X$ is a vector field and $\al$ a $k$-form on a manifold $M$, then the Lie 
derivative of $\al$ with respect to $X$ (i.e. along the flow of $X$) is
\be
{\cal L}_X \al = {\bf d}{\bf i}_X \al + {\bf i}_X{\bf d}\al \;\; .
\label{magic}
\ee
This is proved by straightforward calculation.

\subsection{Symplectic manifolds; the cotangent bundle as a symplectic 
manifold}\label{cotgtblesymp}
Any cotangent bundle $T^*Q$ has a natural  {\em symplectic structure}, which is 
the geometric structure on manifolds corresponding to the symplectic matrix $\o$ 
introduced by eq. \ref{eq;defineOmega}, and to the symplectic forms on vector 
spaces defined at the end of Section \ref{formsassoclin}. (Here `natural' means 
intrinsic, and in particular, independent  of a choice of coordinates or bases.) 
It is this structure  that enables a scalar function to determine a dynamics. That 
is: the symplectic  structure implies that any scalar function $H:T^*Q \rightarrow 
\mathR$ defines a vector field $X_H$ on $T^*Q$.\\
\indent I first describe this structure (Section \ref{453A}), and then show that 
any cotangent bundle has it (Section \ref{453B}). Later subsections will develop 
the consequences.

\subsubsection{Symplectic manifolds}\label{453A}
A {\em symplectic structure} or {\em symplectic form} on a manifold $M$ is defined 
to be a differential 2-form $\o$ on $M$ that is closed (i.e. ${\bf d} \o = 0$) and 
non-degenerate. That is: for any $x \in M$, and any two tangent  vectors at $x$, 
$\sigma, \tau \in T_x$:
\be
{\bf d}\o = 0 \;\; \mbox{ and } \;\; \forall \; \tau \neq 0, \;\; \exists \sigma: 
\;\;\; \o(\tau,\sigma) \neq 0 \;\; .
\label{definesympstruc} 
\ee 
Such a pair $(M, \o)$ is called a {\em symplectic  manifold}.

There is a rich theory of symplectic manifolds; but we shall only need a small 
fragment of it, building on our discussion in Section \ref{formsassoclin}. (In 
particular, the fact that we mostly avoid the theory of canonical transformations 
means we will not need the theory of Lagrangian sub-manifolds.)\\
\indent First, it follows from the non-degeneracy of $\o$ that $M$ is 
even-dimensional; (cf.  eq. \ref{eq;gramschmidt}).\\
\indent  It also follows that at any $x \in M$, there is a basis-independent 
isomorphism $\o^{\fl}$ from the tangent space $T_x$ to its dual $T^*_x$. We saw 
this in (2) and (4) of Section \ref{formsassoclin}, especially eq. 
\ref{definebil'sassoclin}. Namely: for any $x \in M$ and $\tau \in T_x$, the value 
of the 1-form $\o^{\fl}(\tau) \in T^*_x$ is defined by
\be
\o^{\fl}(\tau)(\sigma) := \o(\sigma,\tau) \;\;\; \forall \sigma \in T_x \; .
\label{symimpliescanlisomm}
\ee
Here we return to the main idea emphasised already in Section 
\ref{DetsysevolngradH}: that symplectic structure enables a covector field, i.e. a 
differential one-form, to determine a vector field. Thus for any function $H: M 
\rightarrow \mathR$, so that $dH$ is a differential 1-form on $M$, the inverse of 
$\o^{\fl}$ (which we might write as $\o^{\sh}$), carries $dH$ to a vector field on 
$M$, written $X_H$. Cf. eq. \ref{introX_H}.

So far, we have noted some implications of $\o$ being non-degenerate. The other 
part of the definition of a symplectic form (for a manifold), viz. $\o$ being 
closed, ${\bf d} \o = 0$, is also important. We shall see in Section 
\ref{noetcomplete} that it implies that a vector field $X$ on a symplectic 
manifold $M$ preserves the symplectic form $\o$ (i.e.  in more physical jargon: 
generates (a one-parameter family of) canonical transformations) iff $X$ is 
Hamiltonian in the sense of Section \ref{hamnvfs}; i.e. there is a scalar function 
$f$ such that  $X = X_f \equiv \o^{\sh}(df)$. Or in terms of the Poisson bracket, 
with $\cdot$ representing the argument place for a scalar function: $ X(\cdot) = 
X_f(\cdot) \equiv \{\cdot, f\}$.

So much by way of introducing symplectic manifolds. I turn to showing that any 
cotangent bundle $T^*Q$ is such a manifold.

\subsubsection{The cotangent bundle}\label{453B}
Choose any local coordinates $q$ on $Q$ (dim($Q$)=$n$), and  the natural local 
coordinates $q,p$ thereby induced on $T^*Q$; (cf. (B) of Section 
\ref{canlmoma1forms}). We define the 2-form 
\be
dp \wedge dq := dp_i \wedge dq^i := \Sigma^n_{i=1} dp_i \wedge dq^i \; .
\label{defineomega}
\ee
To show that eq. \ref{defineomega} defines the same 2-form, whatever choice we 
make of the chart $q$ on $Q$, it suffices to show that $dp \wedge dq$ is the 
exterior derivative of a 1-form on $T^*Q$ which is  defined naturally (i.e. 
independently of coordinates or bases) from the derivative (also known as: 
tangent) map of the projection
\be
\pi: (q,p) \in T^*Q  \mapsto q \in Q .
\ee 
Thus consider a tangent vector $\tau$ (not to $Q$, but) to the cotangent bundle 
$T^*Q$  at a point $\eta = (q,p) \in T^*Q,$ i.e. $q \in Q$ and $p \in T^*_q$. Let 
us write this as: $\tau \in T_{\eta}(T^*Q) \equiv T_{(q,p)}(T^*Q)$. The derivative 
map, $D \pi$ say,  of the natural projection $\pi$ applies to $\tau$:
\be
D\pi: \tau \in T_{(q,p)}(T^*Q) \mapsto (D\pi(\tau)) \in T_q \; \; .
\label{derivprojectionpi}
\ee
Now define a 1-form $\theta_H$ on $T^*Q$ by
\be
\theta_H: \tau \in  T_{(q,p)}(T^*Q) \mapsto p(D\pi(\tau)) \in \mathR \; ;
\label{definethetaH}
\ee 
where in this definition  of $\theta_H$, $p$ is defined to be the second component 
of $\tau$'s base-point $(q,p) \in T^*Q$; i.e. $\tau \in T_{(q,p)}(T^*Q)$ and $p 
\in T^*_q$.

This 1-form is called the {\em canonical 1-form} on $T^*Q$. It is the 
``Hamiltonian version'' of the 1-form $\theta_L$ defined by eq. \ref{defpdq}; and 
also there called the `canonical 1-form'. But Section \ref{canlmoma1forms}'s 
discussion of the ``fruitful ambiguity'' of the symbol $p$ brings out a contrast.  
While $\theta_L$ as defined by eq. \ref{defpdq} clearly depends on $L$, the 
definition of $\theta_H$, eq. \ref{definethetaH}, does {\em not} depend on any 
function $H$. $\theta_H$ is given just by the cotangent bundle structure. Hence 
the subscript $H$ here just indicates ``Hamiltonian (as against Lagrangian) 
version'', {\em not} dependence on a function $H$.

\indent So much by way of a natural definition of a 1-form. One now checks that in 
any natural local coordinates $q,p$, $\theta_H$ is given by 
\be
\theta_H = p_i dq^i.
\label{definethetaHagain}
\ee
Finally, we define a 2-form by taking the exterior derivative of $\theta_H$:
\be
 {\bf d}(\theta_H) := {\bf d}(p_i dq^i) \equiv dp_i \wedge dq^i \; .
\label{defineDthetaH}
\ee
where the last equation follows immediately from eq. \ref{extderivkformincooords}. 
One checks that this 2-form is closed (since ${\bf d}^2 = 0$) and non-degenerate. 
So $(T^*Q, {\bf d}(\theta_H))$ is a symplectic manifold.

Referring to eq. \ref{explainARn} of Section \ref{sympformintro}, or eq. 
\ref{defcanlsympformvecsp}  of Section \ref{formsassoclin}, or eq. 
\ref{explainARnagain}  of Section \ref{introduceforms}, we see that at each point 
$(q,p) \in T^*Q$, this symplectic form is, upto a sign, our familiar ``sum of 
signed areas''---first seen as induced by the matrix $\o$ of eq. 
\ref{eq;defineOmega}. 

Accordingly, Section \ref{formsassoclin}'s definition of a canonical symplectic 
form is extended to the present case: ${\bf d}(\theta_H)$, or its negative $-{\bf 
d}(\theta_H)$, is called the {\em canonical symplectic form}, or {\em canonical 
2-form}. (The difference from Section \ref{formsassoclin}'s definition is that on 
a manifold, the symplectic form is required to be closed.)

\indent (The difference by a sign is of course conventional: it arises from our 
taking the $q$s, not the $p$s, as the first $n$ out of the $2n$ coordinates. For 
if we had instead taken the $p$s, the matrix occurring in eq. 
\ref{eq;hamwithOmatrixnotation} would have been $- \o \equiv \o^{-1}$: exactly 
matching the cotangent bundle's intrinsic 2-form ${\bf d}(\theta_H)$.) 

We will see, in Section \ref{darboux}, a theorem (Darboux's theorem) to the effect 
that locally, any symplectic manifold ``looks like'' a cotangent bundle: or in 
other words,   a cotangent bundle is locally a ``universal'' example of  
symplectic structure. But first we return, in the next two Subsections, to 
Hamilton's equations, and Noether's theorem.  

\subsection{Geometric formulations of Hamilton's equations}\label{geomHamEq}
We already emphasised in Sections \ref{sympformintro} and \ref{PoissNoet} the main 
geometric idea behind Hamilton's equations: that a gradient, i.e. covector, field 
$dH$ determines a vector field $X_H$. We first saw this determination  via the 
symplectic matrix, in eq. \ref{introX_H} of Section \ref{DetsysevolngradH}, viz.
\be
X_H (z) = \o \nabla H(z) \; ;
\label{introX_Hrepeat}
\ee  
and then via the Poisson bracket, in eq. \ref{HamDelta0} of Section \ref{hamnvfs}, 
viz.
\be
D := X_H = \frac{d}{dt} = {\dot q}^i \frac{\pl }{\pl q^i} + {\dot p}_i \frac{\pl 
}{\pl p_i} = 
\frac{\pl H}{\pl p_i} \frac{\pl }{\pl q^i} - \frac{\pl H}{\pl q^i} \frac{\pl }{\pl 
p_i} = \{\cdot, H\}
\;.
\label{HamDelta0repeat}
\ee   
The symplectic structure and Poisson bracket were related by eq. \ref{pbasmatrix}, 
viz.
\be
\{f, g\} (z) = {\tilde {\nabla f}}(z). \o. \nabla g (z).
\label{pbasmatrixrepeat}
\ee
And to this earlier discussion, the last Subsection, Section \ref{cotgtblesymp}, 
added the identification of the canonical symplectic form of a cotangent bundle, 
eq. \ref{defineDthetaH}. 

Let us sum up these discussions by giving some geometric formulations of 
Hamilton's equations at a point $z = (q,p)$ in a cotangent bundle $T^*Q$. Let us 
write $\o^{\sh}$ for the (basis-independent) isomorphism from the cotangent space 
to the tangent space, $T^*_z \rightarrow T_z$, induced by $\o := - {\bf 
d}(\theta_H) = dq^i \wedge dp_i$ (cf. eq. \ref{othersharpasinverse} and 
\ref{symimpliescanlisomm}). Then Hamilton's equations, eq. \ref{introX_H} or 
\ref{introX_Hrepeat}, may be written as:
\be
{\dot z} = X_H (z) = \o^{\sh} ({\bf d}H(z)) = \o^{\sh} (dH(z)) \;\; .
\label{HEgeomic1}
\ee
Applying $\o^{\fl}$, the inverse isomorphism $T_z \rightarrow T^*_z$, to  both 
sides, we get
\be
\o^{\fl} X_H (z) = dH(z) \;\; .
\label{HEgeomic2}
\ee
In terms of the symplectic form $\o$ at $z$, this is (cf. eq. 
\ref{definebil'sassoclin}): for all vectors $\tau \in T_z$
\be
\o(X_H(z), \tau) = dH(z) \cdot \tau \;\; ;
\label{HEgeomic3}
\ee
or in terms of the contraction defined by eq. \ref{definecontraction}, with 
$\cdot$ marking the argument place of $\tau \in T_z$:
\be
{\bf i}_{X_H} \o :=  \o(X_H(z), \cdot) = dH(z)(\cdot) \;\; .
\label{HEgeomic4}
\ee
More briefly, and now for any function $f$, it is:
\be
{\bf i}_{X_f} \o = df \; .
\label{HEgeomic4forf}
\ee

Here is a final example. Recall the relation between the Poisson bracket and the 
directional derivative  (or the Lie derivative  $\cal L$) of a function, eq. 
\ref{defineX_fusual} and \ref{HamDelta0repeat}: viz.
\be
{\cal L}_{X_f} g = dg(X_f) = X_f(g) = \{ g, f \} \; .
\label{Liepbbasic}
\ee
Combining this with eq. \ref{HEgeomic4forf}, we can reformulate the relation  
between the symplectic form and Poisson bracket, eq. \ref{pbasmatrixrepeat}, in 
the form: 
\be
\{g, f \} = dg(X_f) = {\bf i}_{X_f} dg = {\bf i}_{X_f} ({\bf i}_{X_g} \o) = 
\o(X_g,X_f) \; .
\label{sympPBgeomic}
\ee

\subsection{Noether's theorem completed}\label{noetcomplete}
 The discussion of Noether's theorem in Section \ref{noetHamsimple} left 
unfinished business: to prove that a vector field generates a one-parameter family 
of canonical transformations iff it is a Hamiltonian vector field (and so justify 
the third claim of Section \ref{443A}). Cartan's magic formula and the 
Poincar\'{e} Lemma, both from Section \ref{introduceforms}, make it easy to prove 
this, for a vector field on any symplectic manifold $(M,\o)$. ($(M,\o)$ need not 
be a cotangent bundle.) 

 We define a vector field $X$ on a symplectic manifold $(M,\o)$ to be {\em 
symplectic} (also known as: {\em canonical}) iff  the Lie-derivative along $X$ of  
the symplectic form vanishes, i.e. ${\cal L}_X \o = 0$.\footnote{As announced in 
Section \ref{limitsgeomperspLag}, I assume the notion of the Lie-derivative, in 
particular the Lie-derivative of a 2-form. Suffice it to say, as a sketch, that 
the flow of $X$ defines a  map on $M$ which induces a map on curves, and so on 
vectors, and so on co-vectors, and so on 2-forms such as $\omega$. Nor will I go 
into details about   the equivalence between this definition of $X$'s being 
symplectic, and $X$'s generating (active) canonical transformations, or preserving 
the Poisson bracket. For as I have emphasised, I will not need to develop the 
theory of canonical transformations.} 

Since $\o$ is closed, i.e. ${\bf d}\o = 0$, Cartan's magic formula, eq. 
\ref{magic}, applied to $\o$ becomes
\be
{\cal L}_X \o \equiv {\bf d}{\bf i}_X \o + {\bf i}_X{\bf d}\o  =  {\bf d}{\bf i}_X 
\o \;\; .
\label{magicforomega}
\ee
So for $X$ to be symplectic is for ${\bf i}_X \o$ to be closed. But by the 
Poincar\'{e} Lemma, if ${\bf i}_X \o$ is closed, it is locally exact. That is: 
there locally exists a scalar function $f: M \rightarrow \mathR$ such that 
\be
{\bf i}_X \o = df \;\; {\rm{i.e.}} \;\; X = X_f \; .
\ee
So for $X$ to be symplectic is equivalent to $X$ being {\em locally Hamiltonian}. 

So we can sum up Noether's theorem from a geometric perspective, as follows.
We define a {\em Hamilton system} to be a triple $(M, \o, H)$ where $(M,\o)$ is a 
symplectic manifold and $H: M \rightarrow \mathR$, i.e. $M \in {\cal F}(M)$.
 We define a (continuous) {\em symmetry} of a Hamiltonian  system to be a vector 
field $X$ on $M$ that preserves both the symplectic form, ${\cal L}_X \o = 0$, and 
the Hamiltonian function, ${\cal L}_X H = 0$. As we have just seen: for any  
symmetry so defined, there locally exists an $f$ such that $X = X_f$. So we can 
apply the ``one-liner'', eq. \ref{naivenoetham}, i.e. the antisymmetry of the 
Poisson bracket,
\be
X_f(H) \equiv \{H, f \} = 0 \;\;\; {\rm{iff}} \;\;\; X_H(f) \equiv \{f, H \} = 0 
\; ,
\ee 
to conclude that $f$ is a first integral (constant of the motion). Thus we have
\begin{quote}
{\bf {Noether's theorem for a Hamilton system}} If $X$ is a symmetry of a 
Hamiltonian system $(M, \o, H)$, then locally $X = X_f$ and $f$ is a constant of 
the motion. And conversely: if $f: M \rightarrow \mathR$ is a constant of the 
motion, then $X_f$ is a symmetry. Besides, this result encompasses the Lagrangian  
version of the theorem; cf. Sections \ref{Noetsubsubsec}  and \ref{noetHamsimple}. 
\end{quote}

Example:--- For most Hamiltonian systems in euclidean space $\mathR^3$, spatial 
translations and rotations are (continuous) symmetries.  For example, consider  
$N$ point-particles interacting by Newtonian gravity. The Hamiltonian is a sum of 
two terms, which are each individually invariant under these euclidean motions:\\
\indent  (i) a kinetic energy term $K$; though I will not go into details, it is 
in fact defined by the euclidean metric of $\mathR^3$ (cf. footnote 4 in Section 
\ref{Lageqsec}), and is thereby invariant; and\\
\indent  (ii) a potential energy term $V$; it depends only on the particles' 
relative distances, and is thereby invariant.\\
\indent The corresponding conserved quantities are the total linear and angular 
momentum.\footnote{By the way, this Hamiltonian is {\em not} invariant under 
boosts. But as I said in Section \ref{limitsgeomperspLag} and footnote 8, I 
restrict myself to time-independent transformations; the treatment of symmetries 
that ``represent the relativity of motion'' needs separate discussion.}

Finally, an incidental remark which relates to the ``rectification theorem'', that 
on any manifold any vector field $X$ can be ``straightened out'' in a 
neighbourhood  around any point at which $X$ is non-zero, so as to have all but 
one component vanish and the last component  equal to 1; cf. eq. 
\ref{arnold10deparallelized}.  Using this theorem, it is easy to see that on any 
even-dimensional manifold {\em any} vector field $X$ is locally Hamiltonian, with 
respect to {\em some} symplectic form, around a point where $X$ is non-zero. (One 
defines the symplectic form by Lie-dragging from a surface transverse to $X$'s 
integral curves.)

\subsection{Darboux's theorem, and its role in reduction}\label{darboux}
{\em Darboux's theorem} states that cotangent bundles are, locally, a ``universal 
form'' of symplectic manifold. That is: Not only is any symplectic manifold ($M, 
\o$) even-dimensional. Also, it ``looks locally like'' a cotangent bundle, in that 
around any $x$ in $M$, there is a local coordinate system $(q^1,...,q^n; 
p_1,...,p_n)$---where the use of both upper and lower indices is now just 
conventional, with no meaning about dual bases!---in which:\\
\indent (i) $\o$ takes the form $dq^i \wedge dp_i$; and so\\
\indent (ii) the Poisson brackets of the $q$s and $p$s take the fundamental form 
in eq. \ref{fundlPBs}.\\
(The theorem generalizes to the Poisson manifolds mentioned in Section 
\ref{Glimpsegenl}.)

\indent Besides, the proof of Darboux's theorem yields further information: 
information  which is important for reducing problems. It arises from the 
beginning of the proof; and will return us to Section \ref{Hameq}'s point that the 
elementary connection between cyclic coordinates and conserved conjugate momenta 
underpins the role of symmetries and conserved  quantities in reductions on 
symplectic manifolds.\\
\indent (In fact, Darboux's theorem also yields two other broad implications about 
reducing problems; but I will not develop the details here. The second implication 
concerns the way that a Hamiltonian structure is preserved in the reduced problem. 
The third implication concerns the requirement that constants of the motion be in 
involution, i.e. have vanishing Poisson bracket with each other; so it leads to 
the idea of complete integrability---a topic this paper foreswears.) 

Namely, the proof implies that ``almost'' any scalar function $f \in {\cal F}(M)$ 
can be taken as the first ``momentum'' coordinate $p_1$; or as the first 
configurational coordinate $q^1$. Here ``almost'' is not meant in a 
measure-theoretic sense; it is just that $f$ is subject to a mild restriction, 
that $df \neq 0$ at the point  $x \in M$.\\ 
\indent In a bit more detail:  The proof of Darboux's theorem starts by taking any 
such $f$ to be our $p_1$, and then constructs the canonically conjugate 
generalized coordinate $q^1$, i.e. the coordinate such that $\{q^1, p_1\} = 1$: so 
that $p_1$ generates translation in the direction of increasing $q^1$. Indeed the 
construction is geometrically clear. The symplectic structure means that any such 
$f$ defines a Hamiltonian vector field $X_f$, and a flow $\phi^f$. We choose a 
$(2n-1)$-dimensional local submanifold $N$ passing through the given point $x$, 
and transverse to all the integral curves of $X_f$ in a neighbourhood of $x$; and 
we set the parameter $\lambda$ of the flow $\phi^f$ to be zero at all points $y 
\in N$. Then for any $z$ in a suitably small neighbourhood of the given point $x$, 
we define the function $q^1(z)$ to be the parameter-value at $z$ of the integral 
curve of $X_f$ that passes through $z$. So by construction, (i) $f$ generates 
translation in the direction of increasing $q^1$, and (ii) defining $p_1 := f$, we 
have  $\{q^1, p_1\} = 1$.\\
\indent This is just the beginning of the proof. But I will not need details of 
how it goes on to establish the local existence of canonical coordinates, i.e. 
coordinates such that analogues of (i) and (ii), also for $i \neq 1$, hold. In 
short, the strategy is to use induction on the dimension of the manifold; for 
details, cf. e.g. Arnold (1989:  230-232).

\indent To see the significance of this for reducing problems, suppose that there 
is a constant of the motion, and that we take it as our $f$, i.e. as the first 
momentum coordinate $p_1$. So the system evolves on a $(2n-1)$-dimensional 
manifold given by an equation $f =$ constant. So writing $H$ in the canonical 
coordinate system secured by Darboux's theorem, we conclude that $0 = {\dot f} 
\equiv - \frac{\pl H}{\pl q^1}$. That is, $q^1$ is cyclic. So as discussed in 
Section \ref{Hameq}, we need only solve the problem in the $2n - 2$ variables 
$q^2,...,q^n; p_2,...,p_n$. Having done so, we can find $q^1$ as a function of 
time, by solving eq. \ref{qnbyquadrat} by quadrature.

To put the point in geometric terms:---\\
\indent (i): The system is confined to a $(2n-1)$-dimensional manifold  $p_1 = \al 
=$ constant, $M_{\al}$ say.\\
\indent (ii): $M_{\al}$ is foliated by a local one-parameter family of 
$(2n-2)$-dimensional manifolds labelled by values of $q^1 \in I \subset \mathR$, 
$M_{\al} = \cup_{q^1 \in I} M_{\al, q^1}$.\\
\indent (iii): Of course, the dynamical vector  field is transverse to the leaves 
of this foliation; i.e. $q^1$ is not a constant of the motion, ${\dot q}^1 \neq 
0$. But since $q^1$ is ignorable, $\frac{\pl H}{\pl q^1} = 0$, the problem to be 
solved is ``the same'' at points $x_1, x_2$ that differ only in their values of 
$q^1$.

\subsection{Geometric formulation of the Legendre 
transformation}\label{geomlegetrsfmn}
Let us round off our development of both Lagrangian and Hamiltonian mechanics, by 
formulating the Legendre transformation as a map from the tangent bundle $TQ$ to 
the cotangent  bundle $T^*Q$. In this formulation, the Legendre transformation is 
often called the {\em fibre derivative}.\\
\indent Again, there is a rich theory to be had here. In part, it relates to the 
topics mentioned in Section \ref{Legtrsfmnvarnal}: (i) the description of a 
function (in the simplest case $f:\mathR \rightarrow \mathR$) by its gradients and 
axis-intercepts, rather than by its arguments and values; (ii) variational 
principles. But I shall not go into details about this theory: since this paper 
emphasises the Hamiltonian framework, a mere glimpse of this theory must suffice. 
(References, additional to those in Section \ref{Legtrsfmnvarnal}, include: 
Abraham and Marsden (1978: Sections 3.6-3.8) and Marsden and Ratiu (1999: Sections 
7.2-7.5, 8.1-8.3).)     
  
Let us return to the Lagrangian framework.
We stressed in Section \ref{geomperspLag} that a scalar on the tangent bundle, the 
Lagrangian $L: TQ \rightarrow \mathR$, ``determines everything'': the dynamical 
vector field $D =: D_L$; and so for given initial $q$ and ${\dot q}$, $L$ 
determines a solution, a trajectory in $TQ$, i.e. $2n$ functions of time $q(t), 
{\dot q}(t)$ with the first $n$ functions determining the latter.\\
\indent  For the Legendre transformation, the fundamental points are that:\\
\indent \indent (1): $L$ also determines at any point $q \in Q$, a preferred map 
$FL_q$ from the tangent space $T_q$  to its dual space $T^*_q$. Besides this 
preferred map:\\
\indent \indent (2): extends trivially to a preferred map from all of $TQ$ to 
$T^*Q$; this is the Legendre transformation, understood geometrically;\\
\indent \indent (3): extends, under some technical conditions (about certain kinds 
of uniqueness, invertibility and smoothness), so as to carry geometric objects of 
various sorts defined on $TQ$ to corresponding objects defined on $T^*Q$, and vice 
versa.\\
\indent So under these conditions, the Legendre transformation (together with its 
inverse) transfers the entire description of the system's motion between the 
Lagrangian and Hamiltonian frameworks.\\
\indent I will explain (1) and (2), but just gesture at (3).

(1): Intuitively, the preferred map $FL_q$ from each tangent space $T_q$ to its 
dual space $T^*_q$ is the transition ${\dot q} \mapsto p$. More precisely: since 
$L$ is a scalar on $TQ$, any choice of local coordinates $q$ on a patch of $Q$, 
together with the  induced local coordinates $q,{\dot q}$ on a patch of $TQ$, 
defines the partial derivatives $\frac{\pl L}{\pl {\dot q}}$. At any point $q$ in 
the domain of the local coordinates, this  defines a preferred map $FL_q$ from the 
tangent space $T_q$ to the dual space $T^*_q$: $FL_q: T_q \rightarrow T^*_q$. 
Namely, a vector $\tau \in T_q$ with components ${\dot q}^i$ in the coordinate 
system $q^i$ on $Q$, i.e. $\tau = {\dot q}^i \frac{\pl}{\pl q^i}$ (think of a 
motion through configuration $q$ with generalized velocity $\tau$) is mapped to 
the 1-form whose components in the dual basis $dq^i$ are $\frac{\pl L}{\pl {\dot 
q}^i}$. That is
\be
FL_q: \tau = {\dot q}^i \frac{\pl}{\pl q^i} \; \in \; T_q \;\;\; \mapsto \;\;\; 
\frac{\pl L}{\pl {\dot q}^i} dq^i \; \in \; T^*_q \;\;.
\label{defLegTrsfmGeomic}
\ee 
One easily checks that because the canonical momenta are a 1-form, this definition 
is, despite appearances, coordinate-independent.

(2): An equivalent definition, manifestly coordinate-independent and given for all 
$q \in Q$, is as follows. Given $L: TQ \rightarrow \mathR$, define $FL: TQ 
\rightarrow T^*Q$, the {\em fibre derivative}, by
\be
\forall q \in Q, \; \forall \sigma,\; \tau \in T_q: \;\;
FL(\sigma) \cdot \tau = \frac{d}{ds}\mid_{s=0} \; L(\sigma + s \tau)
\label{definefibrederiv}
\ee
(We here take $\sigma, \tau$ to encode the identity of the base-point $q$, so that 
we make notation simpler, writing $FL(\sigma)$ rather than $FL((q,\sigma))$ etc.) 
That is: $FL(\sigma) \cdot \tau$ is the derivative of $L$ at $\sigma$, along the 
fibre $T_q$ of the fibre bundle $TQ$, in the direction $\tau$. So $FL$ is 
fibre-preserving: i.e. it maps the fibre $T_q$ of $TQ$ to the fibre $T^*_q$ of 
$T^*Q$. In local coordinates $q,{\dot q}$ on $TQ$, $FL$ is given by:
\be
FL(q^i, {\dot q}^i) = (q^i, \frac{\pl L}{\pl {\dot q}^i}) \;\; ; \;\; 
{\rm{i.e.}} \;\; p_i = \frac{\pl L}{\pl {\dot q}^i} \;\; .
\ee

An important special case involves a free system (i.e. no potential term in the 
Lagrangian) and a configuration manifold $Q$ with a metric $g = g_{ij}$ defined by 
the kinetic energy. (Cf. footnote 4 for the definition of this metric: in short, 
the constraints being scleronomous (i.e. time-independent, cf. Section 
\ref{Lageqsec}), implies that for any coordinate system on $Q$, the kinetic energy 
is a homogeneous quadratic form in the generalized velocities.) The Lagrangian is 
then just the kinetic energy of the metric,
\be
L(q, {\dot q}) \equiv L({\dot q}) := \frac{1}{2}g_{ij}{\dot q}^i{\dot q}^j
\label{lagnfreemetric}
\ee
so that the fibre derivative is given by
\be
FL(\sigma) \cdot \tau = g(\sigma,\tau) = g_{ij}\sigma^i \tau^j \;\; , \;\; 
{\rm{i.e.}} \;\;\;
p_i = g_{ij}{\dot q}^j \; .
\label{legendrefreemetric}
\ee

(3): We can use $FL$ to pull-back to $TQ$ the canonical 1-form $\theta \equiv 
\theta_H$ and symplectic form $\o$ from $T^*Q$ (eq. \ref{definethetaH} and 
\ref{definethetaHagain} with $\o = - {\bf d}\theta$, from Section 
\ref{cotgtblesymp}.B). That is, we can define
\be
\theta_L := (FL)^* \theta_H {\rm{\;\; and \;\;}} \o_L := (FL)^* \o \; \; .
\ee
Since exterior differentiation  $\bf d$ commutes with pull-backs, $\o_L = - {\bf 
d}\theta_L$. Furthermore:\\
\indent (i): As one would hope, $\theta_L$, so defined, is Lagrangian mechanics' 
canonical 1-form, which we already defined in eq. \ref{defpdq} (and which played a 
central role in the Lagrangian version of Noether's theorem).\\
\indent (ii): One can show that $\o_L$ is non-degenerate iff the Hessian condition 
eq. \ref{nonzerohessian} holds. So under this condition, we can analyse Lagrangian 
mechanics in terms of symplectic structure.

Given $L$, we define its energy function $E: TQ \rightarrow \mathR$ by
\be
\forall \; v \equiv (q,\tau) \in TQ, \;\;  E(v) := FL(v) \cdot v - L(v) \;\; ;
\ee
or in coordinates
\be
E(q^i,{\dot q}^i) :=  \frac{\pl L}{\pl {\dot q}^i}{\dot q}^i - L(q^i,{\dot q}^i)
\ee
If $FL$ is a diffeomorphism, we find that $E \circ (FL)^{-1}$ is, as one would 
hope, the Hamiltonian function $H: T^*Q \rightarrow \mathR$ which we already 
defined in eq. \ref{defineHamJS}.\\
\indent And accordingly, if $FL$ is a diffeomorphism, then the derivative of $FL$ 
carries the dynamical vector field $\frac{d}{dt}$ in the Lagrangian description, 
as defined in eq. \ref{LagDelta} (Section \ref{geomperspLag}, (2)), viz.
\be
D_L := {\dot q}^i \frac{\pl}{\pl q^i} + {\ddot q}^i \frac{\pl}{\pl {\dot q}^i} 
\;\; ,
\label{LagDeltarepeat}
\ee
to the Hamiltonian dynamical vector field, viz.
\be
D_H := {\dot q}^i \frac{\pl}{\pl q^i} + {\dot p}_i \frac{\pl}{\pl {p_i}} \;\;.
\label{LagDeltarepeat}
\ee

More generally, one can show if $FL$ is a diffeomorphism, there is a bijective 
correspondence between the various geometric structures used in the Lagrangian and 
Hamiltonian descriptions. For precise statements of this idea, cf. e.g. Abraham 
and Marsden (1978: Theorem 3.6.9) and Marsden and Ratiu (1999: Theorem 7.4.3.), 
and their preceding discussions.

\subsection{Glimpsing the more general framework of Poisson 
manifolds}\label{Glimpsegenl} 
 Recall that Section \ref{pbintro} listed several properties of the Poisson 
bracket, as defined by eq. \ref{naivedefinepb} or \ref{definepbfromxi}. We end by 
briefly describing how the postulation of a bracket that acts on the scalar 
functions $F: M \rightarrow \mathR$ defined on {\em any} manifold $M$, and 
possesses four of Section \ref{pbintro}'s listed properties, provides a sufficient 
framework for mechanics in Hamiltonian style. The bracket is again called a 
`Poisson bracket', and the manifold $M$ equipped with such a bracket is called a 
{\em Poisson manifold}. 

Namely, we require the following four 
properties. The Poisson bracket is to be bilinear; antisymmetric; and to obey the 
Jacobi identity (eq. \ref{Jacobiidentityfirst}) for any real functions $F,G,H$ on 
$M$, i.e.
\be
\{\{F, H \}, G \} + \{\{G, F \}, H \} + \{\{H, G \}, F \}= 0 \;\; ;
\label{Jacobiidentity2nd}
\ee
and to obey Leibniz' rule for products (eq. \ref{Pbofproductfirst}), i.e.
\be
\{F, H \cdot G\} = \{F, H\}\cdot G + H \cdot \{F, G\} \;\; . 
\label{Pbofproduct2nd}
\ee

This generalizes Hamiltonian mechanics: in particular, a Poisson manifold need not 
be a symplectic manifold. The main idea of the extra generality is that the 
antisymmetric bilinear map that gives the geometry of the state space (the 
analogue of Section \ref{sympformintro}'s symplectic form $\o$) can be degenerate. 
So this map can ``have extra zeroes'', as in eq. \ref{omegawiths} and 
\ref{eq;gramschmidt}.  (This map  is induced by the generalized Poisson bracket, 
via an analogue of eq. \ref{pbasmatrix}.) This means  that a Poisson manifold  can 
have {\em odd} dimension; while we saw in Section \ref{formsassoclin} that any 
symplectic vector space is even-dimensional---and so, therefore, is any symplectic 
manifold (Section \ref{453A} and \ref{darboux}). 

On the other hand, the generalized framework has strong connections with the usual 
one.\footnote{Because of these  connections, it is natural to still call the more 
general framework `Hamiltonian'; as is usually done. But of course this is just a 
verbal matter.}  One main connection  is the result that any Poisson  manifold $M$  
is a disjoint union of even-dimensional  manifolds, on each of which $M$'s 
degenerate antisymmetric bilinear form (induced by the generalized Poisson 
bracket) restricts to be non-degenerate; so that there is an orthodox Hamiltonian 
mechanics on each such `symplectic leaf'. Another main connection is  that Section 
\ref{noetHamsimple}'s ``one-liner'' version of Noether's theorem, eq. 
\ref{naivenoetham}, underpins versions of Noether's theorem  for the more general 
framework. 

This generalized framework is important for various reasons; I will just mention 
two.\\
\indent (i): For a system whose orthodox Hamiltonian mechanics on a symplectic 
manifold (dimension $2n$, say) depends on $s$ real parameters, it is sometimes 
natural to consider the corresponding  $(2n + s)$-dimensional space. This is often 
a Poisson manifold; viz., one foliated into an $s$-dimensional family of 
$2n$-dimensional symplectic manifolds. This scenario occurs even for some very 
familiar systems, such as the pivoted rigid body described by Euler's equations.\\
\indent (ii): Poisson manifolds often arise in the theory of symplectic reduction. 
For when you quotient a symplectic manifold by the action of a group (e.g. a group 
of symmetries of a Hamiltonian system in the sense of Section \ref{noetcomplete}), 
you often get a Poisson manifold, rather than a symplectic one. Indeed, the 
pivoted rigid body is itself an example of this.\\
\indent But this generalized framework is a large topic, which we cannot go into: 
as mentioned, Butterfield (2006) is a philosopher's introduction.

For now, we end with a historical point.\footnote{As mentioned in footnote 10, 
Olver (2000) gives many details especially about Lie; e.g. Olver (2000: 374-379, 
427-428). Cf. also Marsden and Ratiu (1999: 336-338, 430-432), and for a full 
history, Hawkins (2000).} It is humbling, but also I hope inspiring, reflection 
about one of classical mechanics' monumental figures. Namely:
a considerable part of the modern theory of Poisson manifolds, including their 
uses for the rigid body and for symplectic reduction, was already contained in Lie 
(1890)!

{\em Acknowledgements}:--- I am grateful to the editors, not least for their 
patience; to audiences in Irvine, Oxford, Princeton and Santa Barbara; and to 
Katherine Brading, Harvey Brown, Hans Halvorson, David Malament, Wayne Myrvold, 
David Wallace, and especially Graeme Segal, for conversations, comments---and 
corrections!

\section{References}
R. Abraham and J. Marsden (1978), {\em Foundations of Mechanics}, second edition: 
Addison-Wesley.

V. Arnold (1973), {\em Ordinary Differential Equations}, MIT Press.

V. Arnold (1989), {\em Mathematical Methods of Classical Mechanics}, Springer, 
(second edition).

G. Belot (2003), `Notes on symmetries', in Brading and Castellani (ed.s) (2003), 
pp. 393-412.

K. Brading and E. Castellani (ed.s) (2003), {\em Symmetry in Physics}, Cambridge 
University Press.

H. Brown and P. Holland (2004), `Simple applications of Noether's first theorem in 
quantum mechanics and electromagnetism`, {\em American Journal of Physics} {\bf 
72} p. 34-39. Available at: http://arxiv.org/abs/quant-ph/0302062 and 
http://philsci-archive.pitt.edu/archive/00000995/

H. Brown and P. Holland (2004a), `Dynamical vs. variational symmetries: 
Understanding Noether's first theorem', {\em Molecular Physics}, {\bf 102}, (11-12
Special Issue), pp. 1133-1139.

J. Butterfield (2004), `Some Aspects of Modality in Analytical mechanics',  in 
{\em Formal Teleology and Causality}, ed. M. St\"{o}ltzner, P. Weingartner, 
Paderborn: Mentis. \\
Available at Los Alamos arXive: http://arxiv.org/abs/physics/0210081 or \\   
http://xxx.soton.ac.uk/abs/physics/0210081;
 and at Pittsburgh archive: http://philsci-archive.pitt.edu/archive/00001192.

J. Butterfield (2004a), `Between Laws and Models: Some Philosophical Morals of 
Lagrangian Mechanics'; available at Los Alamos arXive: 
http://arxiv.org/abs/physics/0409030 or \\   
http://xxx.soton.ac.uk/abs/physics/0409030;
 and at Pittsburgh archive: http://philsci-archive.pitt.edu/archive/00001937/.
 
J. Butterfield (2004b), `On Hamilton-Jacobi Theory as a Classical Root of Theory', 
in A. Elitzur, S. Dolev and N. Kolenda (eds.), {\em Quo Vadis Quantum Mechanics?}, 
Springer, pp. 239-273; available at Los Alamos arXive: 
http://arxiv.org/abs/quant-ph/0210140; or at Pittsburgh archive: 
http://philsci-archive.pitt.edu/archive/00001193/
 
J. Butterfield (2005), `Between Laws and Models: Some Philosophical Morals of 
Hamiltonian Mechanics', in preparation.

J. Butterfield (2006), `On Symplectic Reduction in Classical Mechanics', 
forthcoming in {\em The North Holland Handbook of Philosophy of Physics}, ed. J. 
Earman and J. Butterfield, North Holland. 

R. Courant and D. Hilbert (1953), {\em Methods of Mathematical Physics}, volume I, 
Wiley-Interscience (Wiley Classics 1989).

R. Courant and D. Hilbert (1962), {\em Methods of Mathematical Physics}, volume 
II, Wiley-Interscience (Wiley Classics 1989).

E. Desloge (1982), {\em Classical Mechanics}, John Wiley.

J. Earman (2003), `Tracking down gauge: an ode to the constrained Hamiltonian 
formalism', in Brading and Castellani (ed.s) (2003), pp. 140-162.

H. Goldstein et al. (2002), {\em Classical Mechanics}, Addison-Wesley, (third 
edition). 

T. Hawkins (2000), {\em Emergence of the Theory of Lie Groups: an essay in the 
history of mathematics 1869-1926}, New York: Springer. 

M. Henneaux and C. Teitelboim (1992), {\em Quantization of Gauge Systems}, 
Princeton University Press.

O. Johns (2005), {\em Analytical Mechanics for Relativity and Quantum Mechanics}, 
Oxford University Press, forthcoming.

J. Jos\'{e} and E. Saletan (1998), {\em Classical Dynamics: a Contemporary 
Approach}, Cambridge University Press.

H. Kastrup (1987), `The contributions of Emmy Noether, Felix Klein and Sophus Lie 
to the modern concept of symmetries  in physical  systems', in {\em Symmetries in 
Physics (1600-1980)}, Barcelona: Bellaterra, Universitat Autonoma de Barcelona, p. 
113-163.

S. Lie (1890). {\em Theorie der Transformationsgruppen: zweiter abschnitt}, 
Leipzig: B.G.Teubner.  

C. Lanczos (1986), {\em The Variational Principles of Mechanics}, Dover; (reprint 
of the 4th edition of 1970).

J. Marsden and T. Ratiu (1999), {\em Introduction to Mechanics and Symmetry}, 
second edition: Springer-Verlag.

G. Morandi et al (1990), `The inverse problem of the calculus of variations and 
the geometry of the tangent bundle', {\em Physics Reports} {\bf 188}, p. 147-284.

P. Olver (2000), {\em Applications of Lie Groups to Differential Equations}, 
second edition: Springer-Verlag.

D. Wallace (2003), `Time-dependent Symmetries: the link between gauge symmetries 
and indeterminism', in Brading and Castellani (ed.s) (2003), pp. 163-173.

E. Wigner (1954), `Conservation laws in classical and quantum physics', {\em 
Progress of Theoretical Physics} {\bf 11}, p. 437-440. 

\end{document}